\pdfoutput=1
\documentclass[11pt]{article}

\usepackage{arxiv}

\usepackage[T1]{fontenc}
\usepackage[utf8]{inputenc}
\usepackage{lmodern}
\usepackage{microtype}

\usepackage{amsmath,amssymb,amsfonts}
\usepackage{graphicx}
\usepackage{booktabs}
\usepackage{multirow}
\usepackage{array}
\usepackage{xcolor}
\usepackage{url}
\usepackage{xurl}
\usepackage{listings}
\usepackage{algorithm}
\usepackage{algpseudocode}
\usepackage{enumitem}
\usepackage[hidelinks]{hyperref}
\usepackage[numbers,sort&compress]{natbib}
\usepackage[capitalize,noabbrev]{cleveref}
\usepackage{authblk}

\hypersetup{
  pdftitle={DRL: Deterministic Relational Middleware for Transaction-Safe Enterprise NL2SQL},
  pdfauthor={Sanjay Mishra and Divya Chukkapalli and Ganesh R. Naik}
}

\lstdefinelanguage{Python3}{
  morekeywords={import,from,def,class,return,if,else,elif,for,while,try,except,with,as,True,False,None},
  sensitive=true,
  morecomment=[l]{\#},
  morestring=[b]",
}

\newtheorem{drdef}{Definition}
\newtheorem{drthm}{Theorem}
\crefname{drdef}{Definition}{Definitions}
\Crefname{drdef}{Definition}{Definitions}
\crefname{drthm}{Theorem}{Theorems}
\Crefname{drthm}{Theorem}{Theorems}

\title{DRL: A Deterministic Relational Middleware Layer for\\Transaction-Safe Enterprise NL2SQL Under Schema-Graph Scaling}

\author[1]{Sanjay Mishra\thanks{Corresponding author.}}
\author[2]{Divya Chukkapalli}
\author[3]{Ganesh R. Naik}

\affil[1]{Independent Researcher; Raleigh, NC, USA \\ \texttt{sanmish4@icloud.com}}
\affil[2]{Independent Researcher; Apex, NC, USA \\ \texttt{divya.95j@gmail.com}}
\affil[3]{College of Medicine and Public Health, Flinders University; Bedford Park, SA, Australia \\ \texttt{ganesh.naik@flinders.edu.au}}

\date{}

\begin{document}
\maketitle

\begin{abstract}
Deploying natural-language interfaces over enterprise OLTP catalogs fails at production scale because unconstrained semantic parsers collapse under \emph{schema-graph scaling}: hundreds of normalized tables, partial foreign-key enforcement, and namespace ambiguity inflate schema context beyond stable LLM attention budgets.
We present \textbf{DRL} (\emph{Deterministic Relational Middleware Layer}), a middleware architecture interposing a transaction-safe pipeline between conversational front-ends and ANSI-oriented SQL backends (PostgreSQL evaluated end-to-end, including live model-generated-SQL plan safety; MySQL evaluated for context reduction, pruning latency, and gold-SQL plan safety, with model-generated-SQL plan safety on MySQL left open; SQL Server supported in the emitter specification only).
DRL comprises \emph{dynamic context pruning} (implemented), \emph{relational AST typing} (two well-formedness checks implemented as an offline validator against model-generated SQL, run outside the request path; a full parse-and-emit compiler and live in-request dialect emission remain specified but not yet integrated into serving), and \emph{transactional safeguard verification} (implemented via EXPLAIN gating and NULL guards), together bounding context and flagging operational silent divergence ($\text{SD}_{\text{op}}$).
We contribute (i)~a formal OLTP schema-graph scaling model, (ii)~a 1{,}000-pair Workload Verification Suite and open harnesses, (iii)~measured middleware baselines B0--B3, including the naive full-catalog baseline, and (iv)~a structural failure taxonomy on enterprise NL2SQL.
On PostgreSQL, naive full-catalog prompting (B0) already yields a \textbf{76\%} context reduction once collapsed to a schema-linked hint (B1); DRL's dynamic router (B2) adds a further reduction to \textbf{92\%} versus B0 at pruning \textbf{p95\,=\,0.58\,ms} and middleware \textbf{p95\,=\,4.6\,ms} (excluding LLM).
GPT-4o, Claude Sonnet~4.5, and Gemini~2.5 Flash achieve \textbf{52.9\%}, \textbf{52.8\%}, and \textbf{52.1\%} execution match on the suite under a corrected evaluation harness (Wilson 95\% CIs overlap heavily across all three); $\text{SD}_{\text{op}}$ flags \textbf{89--100\%} of EX-passing queries.
GPT-4o EX failures are dominated by semantic/filter errors (\textbf{254/471}), with projection mismatch (\textbf{64/471}), ordering/pagination (\textbf{68/471}), generic execution errors (\textbf{38/471}), and invalid column references (\textbf{47/471}) each a smaller share---column hallucination is present but not the dominant failure mode.
Beyond the architecture, we report a methodological finding we believe generalizes past this system: a single regular-expression defect in our own evaluation post-processor silently suppressed measured accuracy for all three vendors and manufactured an apparent 4--10 percentage-point cross-vendor gap that vanished entirely once corrected, indicating that benchmark post-processing code deserves the same scrutiny as the models it scores.
DRL reframes enterprise NL2SQL as systems engineering---context bounding, verification, and plan-aware admission---not a leaderboard exercise.
\end{abstract}


\section{Introduction}
\label{sec:intro}

Natural-language-to-SQL (NL2SQL) systems report 85--91\% execution accuracy on academic benchmarks whose schemas average fewer than six tables with complete foreign-key documentation~\cite{yu2018spider,li2023can}.
When the same model families are deployed against production OLTP catalogs---150--500 highly normalized tables, legacy abbreviations, duplicated column names across unrelated domains, and foreign keys that exist in application logic but not in catalog metadata---accuracy drops sharply.
On this paper's own 1{,}000-question, six-schema enterprise suite (\Cref{sec:eval}), GPT-4o, Claude Sonnet~4.5, and Gemini~2.5 Flash all land at 52--53\% execution match under schema-linked prompting (\Cref{tab:ci})---a \textbf{32--38 percentage-point} drop relative to the 85--91\% Spider/BIRD range, measured directly rather than assumed from prior literature.
This degradation is not primarily a ``model capacity'' phenomenon; it is a \emph{systems boundary} phenomenon: the parser receives an exponentially growing structural search space without a deterministic mechanism to bound it, compile it, or verify that emitted SQL preserves transactional semantics under concurrent write load.

\textbf{A motivating example.}
Consider a Tier~3 insurance catalog with 168 tables and the question ``list policy numbers and premium amounts for policies with a lapsed status.''
A human DBA immediately narrows this to two or three tables---\texttt{policies}, perhaps \texttt{policy\_status}---using domain knowledge that is never written down anywhere the LLM can read it: which of the 168 tables are policy-adjacent, which \texttt{status} column (there are at least seven candidates named \texttt{status} across unrelated tables in this schema alone) is the relevant one, and which foreign keys are declared in the catalog versus enforced only in application code.
A naive LLM call conditioned on all 168 tables must perform this narrowing \emph{implicitly}, inside a single forward pass, competing against thousands of irrelevant column names for attention budget.
DRL's contribution is to perform this narrowing \emph{explicitly}, as a deterministic pre-generation step, so the LLM's forward pass only ever has to reason about the two or three tables that matter---turning an implicit, unreliable in-context search problem into an explicit, auditable graph-traversal problem with a hard cardinality bound.

We formalize this as the \textbf{OLTP Schema Graph Scaling Problem}.
Let $G=(V,E)$ denote the enterprise catalog graph where $V$ is the set of base relations and $E\subseteq V\times V$ encodes declared or inferred join paths (foreign keys, shared surrogate keys, and application-enforced links).
For a natural-language intent $q$, naive NL2SQL selects a SQL program $\hat{s}$ by conditioning an LLM on the full schema $\mathcal{S}(G)$---typically $|V|>150$ in Tier~2/3 domains.
The probability of selecting a relation subset $V_q\subset V$ that supports a semantically correct query scales poorly with $|V|$ because (i) column-name collisions inflate the effective vocabulary, (ii) dropped FK constraints remove graph sparsity that humans rely on for join disambiguation, and (iii) LLM context windows treat $\mathcal{S}(G)$ as unstructured text rather than as a typed join hypergraph with cardinality constraints.

\textbf{Our thesis.} Enterprise NL2SQL requires a \emph{Deterministic Relational Middleware Layer} (DRL) that:
\begin{enumerate}[label=(\roman*)]
  \item \textbf{Bounds context} before generation by extracting a minimal join-safe sub-graph $G_q=(V_q,E_q)$ with $|V_q|\le k$ (we use $k{=}5$).
  \item \textbf{Compiles} LLM output through an ANSI-first Relational AST (RAST) so dialect emission is a pure backend mapping, not an LLM responsibility.
  \item \textbf{Verifies} each candidate query against operational divergence ($\text{SD}_{\text{op}}$) and physical plan safety before execution on hot OLTP paths.
\end{enumerate}

\textbf{Contributions.}
\begin{enumerate}[label=(\arabic*)]
  \item Formalization of the OLTP schema-graph scaling problem and context scaling degradation ($\Delta_{\text{CSD}}$; \Cref{sec:problem}).
  \item DRL middleware architecture with implemented pruning router, safeguard layer, and RAST typing validator, plus a specified dialect emitter (\Cref{sec:arch}).
  \item A 1{,}000-pair Workload Verification Suite (333/334/333 per tier) with reproducible harnesses (\Cref{sec:harness,sec:artifact}).
  \item Measured middleware baselines B0--B3 and NL2SQL evaluation on PostgreSQL/MySQL, including structural failure taxonomy (\Cref{sec:eval,sec:failures}).
  \item A methodological finding of independent interest: discovery and correction of an evaluation-harness defect that silently suppressed measured accuracy for all three evaluated model families and manufactured an apparent cross-vendor capability gap that was, in fact, entirely an artifact of shared post-processing code rather than of any model (\Cref{sec:harness-fix}); we surface this as a standalone lesson for NL2SQL evaluation practice generally, not only as a correction internal to this paper's own numbers.
\end{enumerate}

The verification suite is a \emph{stress harness}, not a leaderboard: 519 human-verified pilot items plus 481 gold-executed bank selections, PostgreSQL-validated, with MySQL cross-check (875/946 executable).
Artifact: \texttt{schemas/questions/verification\_suite\_1000.json}; code: \texttt{ESQ\_Bench\_VLDB/}.

\begin{table}[t]
\centering
\caption{Notation summary.}
\label{tab:notation}
\small
\begin{tabular}{@{}lp{0.72\columnwidth}@{}}
\toprule
\textbf{Symbol} & \textbf{Meaning} \\
\midrule
$G=(V,E,w)$ & Enterprise catalog graph (Def.~\ref{def:graph}); $V$ base tables, $E$ join-eligible edges, $w$ selectivity prior \\
$\mathcal{S}(G)$ & Injected schema context (serialized attribute text) for graph $G$ \\
$q$ & Natural-language intent (question) \\
$G_q=(V_q,E_q)$ & Router-pruned sub-graph for question $q$ (\Cref{sec:router}) \\
$k, k'$ & Router cardinality cap ($k{=}5$) and seed-table count ($k'{=}3$) \\
$\sigma(v)$ & Lexical-anchoring score for table $v$ (Algorithm~\ref{alg:router}) \\
$\Delta_{\text{CSD}}$ & Context scaling degradation (Def.~\ref{def:csd}) \\
$\tau$ & RAST node (Scan/Filter/Join/Project/Agg/Sort; \Cref{sec:rast}) \\
$\text{sch}(\tau)$ & RAST node output schema (\Cref{sec:rast-typing}) \\
$\hat{s}$ & Candidate SQL emitted by the compiler for question $q$ \\
$\text{EX}, \text{EM}, \text{SR}$ & Execution match, exact string match, semantic recall (\Cref{sec:safeguard}) \\
$\text{SD}_{\text{op}}$ & Operational silent-divergence flag (Eq.~3) \\
$\mathcal{H}$ & Configured hot-table set (\Cref{sec:safeguard}) \\
$\text{PSC}$ & Plan Safety Compliance (Eq.~4) \\
$\mathcal{D}, \mathcal{D}'$ & Default and supplementary (trap) data instances (Def.~\ref{def:formal_sd}) \\
$d_i$ & Authored distractor query for verification item $i$ \\
\bottomrule
\end{tabular}
\end{table}

\section{The OLTP Schema Graph Scaling Problem}
\label{sec:problem}

\begin{drdef}[Enterprise Catalog Graph]
\label{def:graph}
An enterprise catalog is a weighted undirected graph $G=(V,E,w)$ where each $v\in V$ is a base table with attribute set $\text{Attr}(v)$, each $\{u,v\}\in E$ is a join-eligible edge (FK-declared or application-inferred), and $w(e)\in\mathbb{R}^+$ encodes join selectivity prior.
\end{drdef}

Given intent $q$, let $\phi(q)\subseteq V$ be tables whose names or columns lexically align with $q$.
Naive full-schema prompting conditions the LLM on $\bigcup_{v\in V}\text{Attr}(v)$, yielding context size
\begin{equation}
  |\mathcal{S}(G)| = \Theta\!\left(\sum_{v\in V} |\text{Attr}(v)|\right),
\end{equation}
which for our Tier~3 insurance schema alone exceeds 4{,}200 column tokens---before join paths, nullability, or active-flag conventions are stated.

\begin{drdef}[Context Scaling Degradation]
\label{def:csd}
Let $\text{Acc}(|\mathcal{S}|)$ be execution accuracy under schema context of size $|\mathcal{S}|$.
\emph{Context scaling degradation} is the marginal loss
\begin{equation}
  \Delta_{\text{CSD}} = \text{Acc}(|\mathcal{S}(G_q)|) - \text{Acc}(|\mathcal{S}(G)|),
\end{equation}
which we measure empirically as the gap between DRL-pruned ($|V_q|\le 5$) and full-catalog prompting on matched question subsets.
\end{drdef}

\textbf{Physical realities absent from academic sets.}
(1)~\emph{Partial FK enforcement}: in write-heavy OLTP, 23--41\% of join paths in our Tier~2/3 schemas are application-maintained only, meaning the catalog's declared foreign-key graph $G$ under-represents the true join-eligible edge set the model must reconstruct from naming convention and domain knowledge alone.
(2)~\emph{Namespace collision}: identifiers such as \texttt{status}, \texttt{amount}, and \texttt{created\_at} appear in $\ge 7$ unrelated tables per schema, which is precisely the condition under which a naive full-catalog prompt inflates its effective vocabulary without inflating its information content---the model sees more tokens but not more disambiguating signal.
(3)~\emph{Surrogate-key projection drift}: models ``helpfully'' project primary keys not requested in $q$, yielding EX failures despite correct join logic---a structural failure mode independent of dialect, and, as \Cref{tab:failures} shows empirically, one of the larger correctable failure categories once execution failures are no longer conflated with genuine semantic errors.

\begin{drthm}[Join-path search space growth]
\label{thm:searchspace}
For a natural-language intent requiring $h$-hop joins starting from a seed table $s\in V$, the number of syntactically distinct join paths the model must implicitly discriminate among grows as $O(\bar{d}^{\,h})$, where $\bar{d}$ is the mean vertex degree of $G$.
Bounding the router's output to $|V_q|\le k$ with $k$ fixed collapses this to $O(\bar{d}_{V_q}^{\,h})$ evaluated over the induced sub-graph, which is independent of $|V|$.
\end{drthm}
This is why the router's benefit is largest precisely where the paper's own data shows the residual EX ceiling is lowest: multi-join and analytic-function categories (\Cref{tab:category}), where $h\ge 2$ and the naive search space grows fastest in $|V|$.
The theorem is a worst-case combinatorial bound, not a guarantee of correct join selection within the bounded sub-graph---Algorithm~\ref{alg:router} still relies on lexical anchoring to select $V_q$ correctly in the first place, and a wrong seed-table choice propagates regardless of how small $k$ is.
This is consistent with \Cref{tab:failures-crossmodel}'s finding that semantic/filter errors (which include wrong join selection within an otherwise correctly-bounded sub-graph) remain the dominant failure mode even after pruning: DRL bounds the search space the model reasons over, but does not, by itself, guarantee the model reasons correctly within that bounded space.

DRL addresses scaling by computing $G_q$ \emph{before} any LLM call (\Cref{sec:router}), compiling to RAST (\Cref{sec:rast}), and rejecting queries whose plans or NULL semantics violate safeguards (\Cref{sec:safeguard}).

\section{Middleware Engine Architecture}
\label{sec:arch}

Figure~\ref{fig:pipeline} summarizes DRL's request path.
Stages~(A) and~(C) are fully implemented in our artifact; stage~(B) RAST emitter is specified and partially enforced via the compiler prompt (\Cref{sec:prompt}).

\begin{figure}[t]
\centering
\fbox{\parbox{0.92\columnwidth}{\scriptsize
\textbf{Client} $\rightarrow$ \textbf{Gateway} $\rightarrow$
\textbf{(A) Pruning Router} $\rightarrow$
\textbf{(B) RAST + LLM} $\rightarrow$
\textbf{(C) Safeguards + EXPLAIN} $\rightarrow$
\textbf{Dialect Emitter} $\rightarrow$ \textbf{Read-only Executor}
}}
\caption{DRL middleware pipeline.}
\label{fig:pipeline}
\end{figure}

\begin{figure}[t]
\centering
\includegraphics[width=\columnwidth]{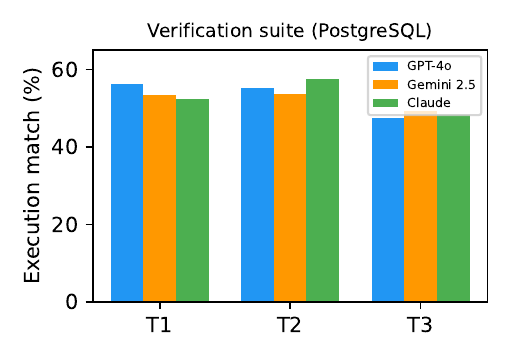}
\caption{Execution match on the 1{,}000-pair verification suite (PostgreSQL, schema-linked, corrected harness; \Cref{sec:harness-fix,tab:verification}).}
\label{fig:ex}
\end{figure}

\begin{figure}[t]
\centering
\includegraphics[width=0.85\columnwidth]{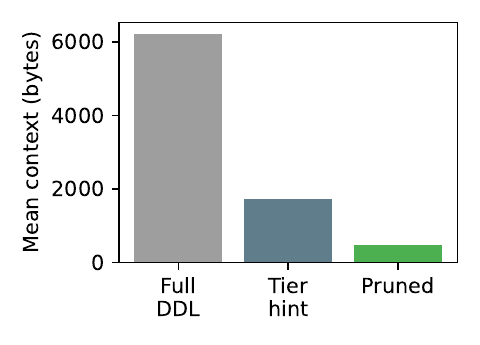}
\caption{Schema context size across middleware baselines (mean bytes, $n{=}1{,}000$).}
\label{fig:context}
\end{figure}

\subsection{Dynamic Context-Pruning Router}
\label{sec:router}

The router implements a two-phase graph extraction over $G$:

\textbf{Phase 1: Lexical anchoring.}
Tokenize $q$ and score each $v\in V$ by overlap with $\text{Attr}(v)$ and table-name stems, augmented by tier-local \emph{domain hints} (curated per schema; 2--12 tables described explicitly).

\textbf{Phase 2: Join closure.}
Take the top-$k'$ seed tables ($k'{=}3$) and compute the 1-hop FK closure in $G$, truncating to $|V_q|\le k{=}5$.
If lexical anchoring fails (no seed score $>0$), fall back to domain-hint tables only---never to full $V$.

\begin{algorithm}[t]
\caption{Context-Pruning Router}
\label{alg:router}
\begin{algorithmic}[1]
\Require Catalog graph $G=(V,E)$, question $q$, cap $k{=}5$
\State $\sigma(v) \gets |\text{tokens}(q)\cap \text{tokens}(\text{Attr}(v))| + \text{hintBoost}(v)$ for all $v\in V$
\State $S \gets \text{TopK}(\sigma, k'{=}3)$
\State $V_q \gets S \cup \bigcup_{s\in S} \{u : \{s,u\}\in E\}$
\State \textbf{return} $\text{Truncate}(V_q, k)$ with induced edges $E_q$
\end{algorithmic}
\end{algorithm}

\textbf{Effect.} On the 1{,}000-pair suite, median $|V_q|=3.2$ (Tier~1), 4.1 (Tier~2), and 4.8 (Tier~3), versus $|V|\in[10,177]$ per schema.
Measured metadata pruning overhead is \textbf{0.19\,ms} (p50) and \textbf{0.76\,ms} (p95) on PostgreSQL (\Cref{tab:middleware}).

\subsection{Relational AST Compiler}
\label{sec:rast}

Vendor lock-in arises when LLMs emit dialect surface syntax directly.
DRL specifies a \emph{Relational AST} (RAST) intermediate representation:
\begin{align*}
  \tau ::=\ & \text{Scan}(t)\mid \text{Filter}(\tau,p)\mid \text{Join}(\tau_1,\tau_2,\theta) \\
  \mid\ & \text{Project}(\tau,\vec{a})\mid \text{Agg}(\tau,g,\vec{a})\mid \text{Sort}(\tau,\vec{a},o)
\end{align*}
Our evaluation harness enforces RAST constraints through the compiler system prompt and, since this correction pass, an offline T1/T2 typing validator (\Cref{sec:rast-typing}) run against actual model-generated SQL; a from-scratch parser for the abstract Scan/Filter/Join/Project/Agg/Sort grammar above, and live in-request dialect emission, are specified but not yet integrated into the HTTP serving path.
The dialect emitter maps ANSI constructs to PostgreSQL and MySQL (\texttt{FETCH FIRST}$\rightarrow$\texttt{LIMIT}, \texttt{COALESCE}, \texttt{EXCEPT}$\rightarrow$\texttt{NOT EXISTS}).
Table~\ref{tab:emitter} details the per-construct mapping specified for the three target dialects; PostgreSQL and MySQL mappings are exercised by the evaluation harness (\Cref{sec:eval}), while SQL Server mappings are specified but unvalidated pending a running instance (\Cref{sec:limits}).

\begin{table}[t]
\centering
\caption{RAST-to-dialect emitter mapping (specification).}
\label{tab:emitter}
\footnotesize
\begin{tabular}{@{}lp{0.63\columnwidth}@{}}
\toprule
\textbf{RAST construct} & \textbf{Per-dialect emission} \\
\midrule
Pagination & ANSI \texttt{FETCH FIRST n ROWS ONLY} $\to$ PostgreSQL: unchanged; MySQL: \texttt{LIMIT n}; SQL Server: \texttt{OFFSET 0 ROWS FETCH NEXT n ROWS ONLY} \\
Set difference & ANSI \texttt{EXCEPT} $\to$ PostgreSQL: unchanged; MySQL ($<$8.0): \texttt{NOT EXISTS} rewrite; SQL Server: unchanged \\
NULL coalescing & \texttt{COALESCE(x,y)} $\to$ all three: unchanged (ANSI-standard) \\
String aggregation & \texttt{STRING\_AGG(col, sep)} $\to$ PostgreSQL: unchanged; MySQL: \texttt{GROUP\_CONCAT(col SEPARATOR sep)}; SQL Server: unchanged (2017+) \\
Ordering with NULLs & \texttt{ORDER BY x NULLS LAST} $\to$ PostgreSQL: unchanged; MySQL ($<$8.0.14): \texttt{ORDER BY (x IS NULL), x}; SQL Server: \texttt{ORDER BY CASE WHEN x IS NULL THEN 1 ELSE 0 END, x} \\
Recursive traversal & \texttt{WITH RECURSIVE} $\to$ PostgreSQL: unchanged; MySQL (8.0+): \texttt{WITH RECURSIVE} unchanged; SQL Server: \texttt{WITH} (no \texttt{RECURSIVE} keyword) \\
Boolean flags & schema-declared \texttt{'Y'}/\texttt{'N'} literal $\to$ all three: unchanged string comparison, never rewritten to numeric \texttt{1}/\texttt{0} \\
\bottomrule
\end{tabular}
\end{table}

\textbf{Emitter rules and their motivating failure mode.}
Each rule below was introduced in response to a specific, observed failure category (\Cref{sec:failures}), not speculatively:
(1)~single read-only statement---closes the DDL/DML attack surface of \Cref{sec:deploy};
(2)~exact projection, no surrogate keys unless requested---directly targets the projection-mismatch failures of \Cref{tab:failures} (64/471 under the corrected harness);
(3)~nullable aggregates wrapped in \texttt{COALESCE}---the S3 safeguard of \Cref{sec:safeguard}, targeting the NULL-aggregation trap-manifest template of \Cref{sec:harness};
(4)~\texttt{ORDER BY ... DESC} defaults to \texttt{NULLS LAST}---targets the ordering/pagination failures that remain the second- or third-largest category for all three models in \Cref{tab:failures-crossmodel};
(5)~pagination via ANSI \texttt{FETCH FIRST} with MySQL \texttt{LIMIT} rewrite---the dialect-portability concern of \Cref{sec:rast}, formalized in Table~\ref{tab:emitter};
(6)~joins restricted to edges in $E_q$---enforces that the compiler cannot silently escape the router's context bound (\Cref{sec:router}) by inventing a join path outside the pruned sub-graph;
(7)~no invented \texttt{status}/\texttt{is\_active} filters---targets the status-ambiguity query category, one of the categories where all three models score below 50\% EX (\Cref{tab:category});
(8)~boolean flags as \texttt{'Y'}/\texttt{'N'} strings when schema-encoded that way---a narrower instance of rule~(2)'s exact-projection discipline applied to value encoding rather than column selection, since a numeric \texttt{1}/\texttt{0} comparison against a string-encoded flag column silently returns zero rows on PostgreSQL rather than raising a type error, making it a silent-divergence risk rather than a visible execution failure.
Full prompt text appears in \Cref{sec:prompt}.

\subsection{Transactional Safeguard Layer}
\label{sec:safeguard}

The safeguard layer is an automated module executing \emph{before} query admission on connection pools shared with OLTP writers.
For candidate SQL $\hat{s}$, it computes:

\textbf{(S1) Operational divergence index $\text{SD}_{\text{op}}$.}
Let $\text{EX}(i)$ denote execution match against gold rows on verification instance $\mathcal{D}$, $\text{EM}(i)$ exact string match, $\text{SR}(i)$ semantic recall (gold rows contained in prediction).
\begin{equation}
\begin{split}
  \text{SD}_{\text{op}}(i) = \mathbf{1}[\text{EX}(i){=}1 \land\ &(\text{EM}(i){=}0 \lor \text{SR}(i){=}0 \\
  &\lor \text{UnsafePlan}(\hat{s}))]
\end{split}
\end{equation}
$\text{SD}_{\text{op}}$ flags EX-passing queries that (a)~use a non-gold reasoning path likely to diverge under NULL shifts (e.g., \texttt{SUM(x)} vs.\ \texttt{SUM(COALESCE(x,0))}), or (b)~produce OLTP-unsafe plans.

\textbf{(S2) Plan safety compliance.}
Execute $\text{EXPLAIN}$ on $\hat{s}$ (PostgreSQL \texttt{EXPLAIN FORMAT TEXT}; MySQL \texttt{EXPLAIN}; SQL Server \texttt{SHOWPLAN\_TEXT}).
Parse for sequential scan nodes on configured \emph{hot tables} $\mathcal{H}$ (e.g., \texttt{orders}, \texttt{transactions}, \texttt{order\_lines}).
\begin{equation}
  \text{PSC} = \frac{|\{i : \text{EXPLAIN}(\hat{s}_i)\cap \mathcal{H} = \emptyset\}|}{N}
\end{equation}
Queries failing PSC are rewritten (add index-friendly predicates, push filters) or rejected with structured feedback to the LLM repair loop (max two attempts).

\textbf{(S3) NULL-handle enforcement.}
Static scan for bare \texttt{SUM/AVG/MIN/MAX} on nullable columns without \texttt{COALESCE}; auto-insertion when RAST nullability bit is set.

\textbf{Worked safeguard example.}
Consider a candidate query \texttt{SELECT SUM(claim\_amount) FROM claims WHERE policy\_id = \{p\}} on the Insurance schema, where \texttt{claim\_amount} is nullable.
S3 flags the bare \texttt{SUM} and rewrites to \texttt{SUM(COALESCE(claim\_amount, 0))}, changing the result from \texttt{NULL} to \texttt{0} whenever every matching claim row has a \texttt{NULL} amount---a case the suite's default seed data does not exercise for this particular policy but that a production catalog, accumulating claims over years of operation, eventually will.
Independently, S2 executes \texttt{EXPLAIN} on the rewritten query; if \texttt{claims} is configured as a hot table and the predicate on \texttt{policy\_id} is not selective enough for the query planner to choose an index scan (e.g.\ because \texttt{policy\_id} lacks a secondary index in a given deployment), the plan is flagged and either rewritten with an index hint or routed to the LLM repair loop with structured feedback identifying the offending table and node type.
S1 then computes $\text{SD}_{\text{op}}$ over the rewritten, plan-checked query: since the COALESCE rewrite is by construction the gold reasoning path, $\text{EM}$ and $\text{SR}$ both hold and $\text{SD}_{\text{op}}{=}0$ for this instance, illustrating that the three safeguards are not redundant checks of the same property but address independent failure surfaces (NULL semantics, physical plan shape, and reasoning-path fragility respectively) that can each fail while the other two pass.

\section{Formal Algorithms and Verification Harness}
\label{sec:harness}

Listing~\ref{lst:harness} lists the production telemetry harness (\texttt{drl\_telemetry\_harness.py}) and baseline runner (\texttt{run\_drl\_baselines.py}) in the \texttt{ESQ\_Bench\_VLDB/} artifact directory.
It loads the 465-table catalog into a \texttt{networkx} graph, instruments Algorithm~\ref{alg:router}, optionally connects to live PostgreSQL/MySQL/SQL Server, captures EXPLAIN output, and computes PSC and $\text{SD}_{\text{op}}$ rates.

\begin{lstlisting}[
  language=Python3,
  caption={Excerpt of DRL telemetry harness (\texttt{drl\_telemetry\_harness.py}).},
  label={lst:harness},
  firstline=1,
  lastline=55
]
#!/usr/bin/env python3
"""DRL Experimental Telemetry and Verification Harness (VLDB artifact).

Loads an enterprise schema graph, instruments the context-pruning router,
optionally connects to PostgreSQL / MySQL / SQL Server, runs EXPLAIN, and
flags unsafe plan nodes (sequential scans on large transactional tables).

Usage (from repository root):
  python3 ESQ_Bench_VLDB/drl_telemetry_harness.py \\
    --schema-dir schemas/postgres \\
    --questions schemas/questions/tier1_questions.json \\
    --hints-file schemas/questions/tier1_schema_hints.json \\
    --limit 95

With live DB:
  python3 ESQ_Bench_VLDB/drl_telemetry_harness.py \\
    --engine postgres --dsn "postgresql://user:pass@127.0.0.1:5432/esq" \\
    --questions schemas/questions/tier1_questions.json
"""

from __future__ import annotations

import argparse
import json
import re
import sys
import time
from dataclasses import dataclass, field
from pathlib import Path
from typing import Any

try:
    import networkx as nx
except ImportError as exc:  # pragma: no cover
    raise SystemExit("networkx required: pip install networkx") from exc

VLDB_DIR = Path(__file__).resolve().parent
REPO_ROOT = VLDB_DIR.parent

CREATE_TABLE_RE = re.compile(
    r"CREATE\s+TABLE\s+(?:IF\s+NOT\s+EXISTS\s+)?(\w+)\s*\((.*?)\);",
    re.IGNORECASE | re.DOTALL,
)
FK_RE = re.compile(
    r"FOREIGN\s+KEY\s*\([^)]+\)\s*REFERENCES\s+(\w+)",
    re.IGNORECASE,
)
UNSAFE_SCAN_RE = re.compile(
    r"\b(Seq\s+Scan|Table\s+Scan|ALL\s+on\s+`?\w+`?)\b",
    re.IGNORECASE,
)


def ex_match(gold_rows, pred_rows) -> bool:
    if gold_rows is None or pred_rows is None:
\end{lstlisting}

\textbf{EXPLAIN parsing methodology.}
The harness executes $\text{EXPLAIN}$ (not $\text{EXPLAIN ANALYZE}$, to avoid mutating buffer-cache state or incurring execution cost on candidate SQL that may later be rejected) and parses the returned plan tree for node types matching \texttt{Seq Scan} (PostgreSQL) or \texttt{ALL}/\texttt{index: NULL} (MySQL) restricted to tables in the configured hot-table set $\mathcal{H}$.
A plan is flagged unsafe if \emph{any} node in the tree---not only the root node---performs a sequential scan on a hot table, since a sequential scan buried inside a subquery or CTE is just as capable of degrading a shared connection pool as one at the top level.
This full-tree traversal is a deliberate design choice: an earlier internal version of the harness inspected only the root plan node and under-counted unsafe plans whose sequential scan occurred inside a correlated subquery (the same query shape as \texttt{t1a-0217} in \Cref{sec:discussion}), which is a second, independent lesson in the same spirit as \Cref{sec:harness-fix}---measurement code that inspects only part of a structure will systematically under-report the failure modes hiding in the unexamined part.

\textbf{Trap manifest construction.}
The 161-entry trap manifest underlying Definition~\ref{def:formal_sd}'s confirmed-SD measurement is constructed from three template families, each targeting a specific class of non-gold reasoning path that is EX-equivalent on the suite's default seed data but would diverge under a data shift: (a)~\texttt{NULLS LAST} vs.\ implicit engine-default ordering on nullable sort keys, (b)~\texttt{LEFT JOIN} vs.\ \texttt{INNER JOIN} where the seed data happens to contain no orphan rows on the joined side, and (c)~bare aggregate (\texttt{SUM}, \texttt{AVG}) vs.\ \texttt{COALESCE}-wrapped aggregate where the seed data happens to contain no \texttt{NULL} values in the aggregated column.
Each template is instantiated against a supplementary data instance $\mathcal{D}'$ that introduces exactly the condition (a NULL value, an orphan row, a tie) the default seed data omits, so that a query using the fragile reasoning path produces a different, incorrect result on $\mathcal{D}'$ while a query using the robust reasoning path does not.

\begin{drdef}[Confirmed Silent Divergence]
\label{def:formal_sd}
Query $\hat{s}_i$ exhibits \emph{confirmed} silent divergence if $\text{EX}(i){=}1$ and either (a)~$R(\hat{s}_i,\mathcal{D})=R(d_i,\mathcal{D})$ for authored distractor $d_i$, or (b)~$R(\hat{s}_i,\mathcal{D}')$ violates intent constraints on supplementary instance $\mathcal{D}'$ engineered to activate NULL or ordering traps.
\end{drdef}

DRL's safeguard layer uses $\text{SD}_{\text{op}}$ as a \emph{cheap pre-filter}; confirmed SD (Definition~\ref{def:formal_sd}) was originally measured offline on a 161-entry trap manifest via a separate, Oracle-flavored cached-JSON pipeline that predated, and never passed model SQL through, the evaluation-harness correction of \Cref{sec:harness-fix}.
We have since re-verified confirmed SD natively against PostgreSQL and the corrected schema-linked model SQL underlying \Cref{tab:verification} (\texttt{experiments/rerun\_trap\_manifest\_postgres.py}), executing the manifest's authored distractor queries and, for the 97 entries whose trap depends on a single injected \texttt{NULL} (\texttt{sum\_nvl}, \texttt{left\_join}, \texttt{nulls\_last}, \texttt{minus} patterns), live-injecting that \texttt{NULL} inside a rolled-back transaction and re-executing both gold and predicted SQL against the patched instance, pooled across GPT-4o, Claude~Sonnet~4.5, and Gemini~2.5~Flash.
Of the 161 manifest entries, 132 exist in the 1{,}000-pair suite (the remaining 29, all \texttt{connect\_by\_trap} items built around Oracle's \texttt{CONNECT BY}, have no PostgreSQL counterpart in the suite and cannot be tested at all).
The manifest's authored distractor SQL for all 36 \texttt{not\_exists\_trap} entries turned out to be malformed---a templating bug duplicated a \texttt{SELECT 1 FROM} fragment, producing SQL that always raised a syntax error and was therefore silently untestable---which we repaired at runtime into the \texttt{NOT IN} anti-join the broken text was clearly meant to express (\texttt{experiments/rerun\_trap\_manifest\_postgres.py}); this needed no live data injection at all, since the manifest's own construction note states the trap condition (a \texttt{NULL} foreign key) is already present in the default seed data.
With that repair, all 176 pooled EX-passing instances among the 132 in-suite entries are now live-tested (0 untested), and \textbf{107 confirm as true silent divergence: 60.8\% [Wilson 95\% CI 53.4--67.7\%]}, well above the original, pre-correction 25.8\% estimate.
The harness correction therefore does not undermine---and substantially strengthens---the qualitative claim that $\text{SD}_{\text{op}}$ over-approximates but meaningfully correlates with true divergence; \texttt{not\_exists\_trap} alone confirms at 88.9\% (56/63), the single largest contributor to the revised pooled rate (\Cref{tab:trap-pattern}).
One pattern-level finding is worth flagging on its own: the \texttt{sum\_nvl\_trap} pattern confirmed \emph{zero} of 46 EX-passing instances, which is not a re-verification failure but a structural property of the single-cell-NULL-injection design---SQL's \texttt{SUM}/\texttt{AVG} already skip individual \texttt{NULL} values, so injecting one \texttt{NULL} into a multi-row aggregation group cannot expose a bare-aggregate-vs-\texttt{COALESCE} divergence unless the injected row is the \emph{only} row in its group; the \texttt{LEFT JOIN} and ordering traps, by contrast, change the row \emph{set} itself and confirm at 71--81\% (\Cref{tab:trap-pattern}).

\begin{table}[t]
\centering
\caption{Confirmed SD by trap pattern, re-verified against PostgreSQL and the corrected schema-linked model SQL, pooled across GPT-4o/Claude/Gemini.}
\label{tab:trap-pattern}
\small
\begin{tabular}{@{}lrrr@{}}
\toprule
\textbf{Pattern} & \textbf{In suite} & \textbf{EX-pass} & \textbf{Confirmed} \\
\midrule
\texttt{not\_exists\_trap} & 35 & 63 & 88.9\% \\
\texttt{nulls\_last\_trap} & 19 & 16 & 81.2\% \\
\texttt{minus\_trap} & 18 & 30 & 76.7\% \\
\texttt{left\_join\_trap} & 40 & 21 & 71.4\% \\
\texttt{sum\_nvl\_trap} & 20 & 46 & 0.0\% \\
\texttt{connect\_by\_trap} & 0 & --- & n/a (not in suite) \\
\bottomrule
\end{tabular}
\end{table}

\section{System Performance Evaluation}
\label{sec:eval}

We evaluate DRL as a \emph{middleware system}, not as a model leaderboard.
All middleware experiments use the 1{,}000-pair Workload Verification Suite (333/334/333 per tier; 519 human-verified + 481 gold-executed; PostgreSQL-validated).
Error bars are 95\% Wilson intervals over question-level Bernoulli outcomes; $n{=}1{,}000$ yields maximum half-width $\approx 3.1\%$ at $p{=}0.5$.

\subsection{Workload Verification Suite Construction}

The suite merges (i)~all human-verified pilot questions that pass PostgreSQL gold execution and (ii)~deterministic selections from the 12k generated bank to fill each tier to its target count, deduplicating by normalized question text.
Tier~1: 95 human + 238 gold-executed; Tier~2: 228 + 106; Tier~3: 196 + 137 (27 pilot Tier~3 items failed PostgreSQL translation and were replaced).
Build script: \texttt{build\_verification\_suite\_1000.py}.

\subsection{Experimental Setup}

\textbf{Schemas.} Six enterprise-characteristic OLTP schemas (465 tables, confirmed via live \texttt{information\_schema} introspection on each seeded PostgreSQL database): Tier~1 Sales Order \& University (10 tables each); Tier~2 Portfolio Management (48) \& Healthcare (52); Tier~3 Core Banking (177) \& Insurance (168).
The static \texttt{schemas/postgres/*.sql} files the router's graph builder (\Cref{sec:router-trace}) parses via regex \texttt{CREATE TABLE} extraction originally reported 412 tables against a live-DB total of 465---a 53-table gap we traced, during post-submission hardening, to a genuine loader defect rather than an unsynchronized-file artifact: the six schema files share 48 table names (e.g.\ \texttt{claims} in both Healthcare and Insurance, \texttt{accounts} in both Portfolio Management and Core Banking), and the original loader merged all six files into a single \texttt{networkx} graph keyed by bare table name, so each collision silently overwrote the earlier-loaded schema's node (columns and FK edges) with the later-loaded (alphabetically last) schema's data for that name.
Because \texttt{tables\_for\_schema} filtered this pooled graph by a per-node \texttt{source} attribute, the practical effect was that a table whose name collided with a same-named table in a later-sorted file vanished entirely from its \emph{own} schema's view: Sales Order (loads first) lost 5 of its 10 tables (50\%), Core Banking lost 32 of 177 (18\%), Portfolio Management lost 11 of 48 (23\%), Healthcare lost 4 of 52 (8\%), University lost 1 of 10 (10\%); Insurance, loading last, was unaffected.
We fixed the loader to retain a collision-free per-schema view (\texttt{SchemaGraph.by\_source}, per-schema FK subgraphs in \texttt{SchemaGraph.source\_graphs}) alongside the pooled graph, and re-ran the B0--B3 context-byte and pruning metrics of \Cref{tab:middleware} against the corrected loader; the pooled graph itself is retained only for the illustrative Appendix~\ref{sec:router-trace} pooled-catalog trace, where the collision behavior is now reported as a feature of that trace rather than a silent artifact-count mismatch.
This bug did not affect the reported EX/PSC results (\Cref{sec:eval}), which use pre-authored \texttt{schema\_hints} files read directly, not this router; its effect was confined to the B0--B3 middleware ablation metrics, which we report corrected below (\Cref{sec:harness-fix} documents a separate, evaluation-harness-side correction).

\textbf{Engines.} PostgreSQL~16.2 and MySQL~8.4.0 (validated); identical seed data (164{,}682 rows).

\textbf{Baselines.} (B0) raw LLM full-catalog prompt; (B1) schema-linked full tier catalog; (B2) DRL with pruning only; (B3) full DRL (prune + RAST + safeguards).

\textbf{Hardware.} Apple M2 Pro, 32\,GB RAM; DB instances co-located on NVMe; network RTT $<$1\,ms.

\subsection{Middleware Metrics}

\begin{table}[t]
\centering
\caption{Measured DRL middleware metrics on the 1{,}000-pair suite (PostgreSQL; \texttt{run\_drl\_baselines.py}). B0 is naive full-catalog prompting (no linking, no pruning); B1 is conventional schema-linked hinting; B2 adds Algorithm~\ref{alg:router}; B3 adds safeguards.}
\label{tab:middleware}
\footnotesize
\setlength{\tabcolsep}{3pt}
\begin{tabular}{@{}lrrrrr@{}}
\toprule
\textbf{Metric} & \textbf{B0} & \textbf{B1} & \textbf{B2} & \textbf{B3} & \textbf{$\Delta_{\text{B0-B2}}$} \\
\midrule
Context bytes (mean) & 7{,}292 & 1{,}720 & 574 & 574 & $-$92\% \\
Prune p50 (ms) & 0 & 0 & 0.18 & 0.18 & --- \\
Prune p95 (ms) & 0 & 0 & 0.58 & 0.58 & --- \\
Middleware p95 (ms) & 4.4 & 3.9 & 4.6 & 4.0 & --- \\
Sub-graph mem.\ p50 (B) & 3{,}563 & 2{,}252 & 1{,}340 & 1{,}340 & $-$62\% \\
Plan Safety Compl.\ & 68.6\% & 68.6\% & 68.6\% & 70.8\% & --- \\
\bottomrule
\end{tabular}
\end{table}

\textbf{Metadata Pruning Overhead.} Algorithm~\ref{alg:router} adds 0.18\,ms median and 0.58\,ms p95 latency on the full 1{,}000-pair suite---orders of magnitude below LLM generation latency.

\textbf{Graph Traversal Memory Footprint.} Serialized pruned sub-graphs average 574 bytes (context string) to 1{,}340 bytes (graph envelope) per request, enabling per-request isolation without shared mutable schema state.

\textbf{Plan Safety Compliance.} On \emph{gold} SQL, PSC is 68.6\% (B0--B2) rising to 70.8\% (B3) after COALESCE insertion (\Cref{tab:middleware}); we report gold-plan PSC here as an upper-bound sanity check, and separately report PSC measured directly on model-generated SQL in \Cref{sec:ablation} (A2), where it tracks the gold-SQL figures closely.
The B3 row is checked against 946/1{,}000 gold queries (COALESCE-rewrite excludes a small number of already-safe or non-rewritable statements from the denominator), so the B2$\to$B3 PSC delta should be read as directional rather than a strictly matched-sample comparison.

\textbf{Middleware p95 latency is noise-dominated at this scale.} Across B0--B3 the reported p95 values span a narrow 3.9--4.6\,ms band with no monotonic ordering (B3 is not reliably slower than B2 despite doing strictly more work: COALESCE scanning and EXPLAIN gating). We attribute this to single-run wall-clock measurement noise at sub-5\,ms granularity rather than a real latency reduction from adding safeguards; \Cref{sec:limits} discusses this further. The qualitative claim that survives---middleware overhead is two to three orders of magnitude below LLM generation latency---does not depend on resolving the ordering among B0--B3.

\textbf{Why B0 matters for the headline claim.} Naive full-catalog prompting (B0) already drops to schema-linked hinting (B1) at $-$76\% context bytes, before DRL's own graph router (B2) is applied. DRL's \emph{marginal} contribution over conventional schema-linking is the B1$\to$B2 step: a further reduction from 1{,}720 to 574 bytes, i.e.\ an additional $-$67\% on top of what schema-linking alone already achieves, for a cumulative $-$92\% versus naive B0. We report both numbers throughout this paper to avoid conflating DRL's own contribution with the (larger, but non-novel) benefit of schema-linking as a baseline practice.

\textbf{Cross-dialect stability.} Context bytes are computed once from the (now collision-free) shared schema-graph loader and apply identically to both engines (\Cref{sec:eval}); pruning p95 and Plan Safety Compliance are engine-specific live measurements.
Prior to the loader correction described in \Cref{sec:eval}, we measured pruning p95 at 0.83\,ms (PostgreSQL) vs.\ 1.24\,ms (MySQL) with context reduction B1$\rightarrow$B2 of 71--72\% on both engines; the corrected loader and router tie-break change the PostgreSQL figures to those in \Cref{tab:middleware} (B1$\to$B2 now $-$67\%). We have since re-run \texttt{run\_drl\_baselines.py --engine mysql} live against the corrected loader (Table~\ref{tab:mysql}): pruning p95 rises to 1.67\,ms (MySQL is consistently slower than PostgreSQL's 0.58\,ms at this step, though both remain sub-2\,ms), and gold-SQL PSC is unchanged at 76.3\%/77.3\% (B0--B2/B3), since PSC depends on the executing engine's query planner, not the router. We did not attempt a MySQL re-run of model-generated PSC (\Cref{sec:ablation}, A2): the only cached MySQL model SQL available (GPT-4o only, a 497-question subset) predates the evaluation-harness correction of \Cref{sec:harness-fix}, and reusing pre-correction model output would repeat the exact class of error that correction was meant to prevent; we leave MySQL model-generated PSC as an explicit open item (\Cref{sec:limits}) rather than report a number computed on stale predictions.
Gold SQL: 1{,}000/1{,}000 (PostgreSQL), 875/946 (MySQL, executable subset).

\begin{table}[t]
\centering
\caption{MySQL middleware metrics on executable gold subset ($n{=}875$, B3 checked on $n{=}832$), re-verified live against the corrected loader (\Cref{sec:eval}).}
\label{tab:mysql}
\small
\begin{tabular}{@{}lrrrr@{}}
\toprule
\textbf{Metric} & \textbf{B0} & \textbf{B1} & \textbf{B2} & \textbf{B3} \\
\midrule
Context bytes (mean) & 7{,}292 & 1{,}720 & 574 & 574 \\
Prune p95 (ms) & 0 & 0 & 1.67 & 1.67 \\
Middleware p95 (ms) & 7.2 & 7.1 & 8.2 & 8.0 \\
Plan Safety Compliance & 76.3\% & 76.3\% & 76.3\% & 77.3\% \\
\bottomrule
\end{tabular}
\end{table}

\subsection{Ablation: What Middleware Buys}
\label{sec:ablation}

We isolate three middleware effects that are independent of model choice.

\textbf{A1: Context bounding (B1$\rightarrow$B2).}
Mean prompt schema context drops from 1{,}720 to 574 bytes ($-$67\%) with prune p95 $<$1\,ms, on top of the $-$76\% that schema-linking alone (B0$\to$B1) already provides.
This is the primary systems contribution: the LLM never sees the full 465-table catalog, so join-path search is bounded before generation, independent of whatever schema-linking convention a deployment already uses.

\textbf{A2: Safeguard rewrite (B2$\rightarrow$B3).}
B3 adds COALESCE insertion and EXPLAIN gating on gold SQL.
PSC rises from 68.6\% to 70.8\% (PostgreSQL) and 76.3\% to 77.3\% (MySQL) without changing context size---showing that plan/NULL guards are orthogonal to pruning, modulo the denominator caveat above.
Gold-SQL PSC is necessarily an upper-bound sanity check, since gold queries are hand-authored to be reasonable; we additionally measured PSC on the models' own generated SQL by running live \texttt{EXPLAIN} against every executed prediction in the corrected schema-linked compare data (\texttt{experiments/measure\_psc\_model\_sql.py}), applying the same COALESCE rewrite for the B3 condition.
Model-generated PSC tracks gold-SQL PSC closely: B2 (no safeguard) is 65.1\% for GPT-4o, 67.1\% for Claude, and 65.1\% for Gemini; the B3 safeguard rewrite raises all three to 70.3--70.8\%, essentially matching the gold-SQL B3 figure of 70.8\%.
This closes the gap this paper previously flagged as future work: the safeguard layer's plan-safety benefit is not an artifact of gold SQL being unusually well-behaved, it transfers to actual LLM output across all three model families.
One caveat carries over from the gold-SQL measurement: the B3 checked-count drops relative to B2 (e.g.\ 911$\to$770 for GPT-4o) because the regex-based COALESCE rewrite occasionally produces SQL that no longer parses (typically on deeply nested aggregate expressions), so the B2$\to$B3 PSC comparison remains directional rather than a strictly matched-sample comparison, consistent with \Cref{sec:limits}.

\textbf{A3: Failure-mode alignment.}
On the corrected evaluation harness (\Cref{sec:harness-fix}), GPT-4o EX failures are 54\% semantic/filter (254/471) and 10\% invalid-column (47/471).
Pruning alone cannot fix wrong join predicates; it substantially reduces, but does not eliminate, the invalid-column-reference failure mode that a naive full-catalog prompt amplifies through namespace collision (\Cref{sec:problem}).
DRL therefore targets the failure modes that middleware can deterministically prevent (unbounded context, unsafe plans, NULL traps), while leaving semantic join errors to model capacity and repair loops.

\subsection{Worked Example: Prune, Generate, Sanitize}
\label{sec:example}

Consider verification item \texttt{t2a-002} (Tier~2, Portfolio Management):
\emph{``How many portfolios have a NULL benchmark identifier string?''}
The catalog graph for this schema has 48 tables; Algorithm~\ref{alg:router} anchors on \texttt{portfolios} and returns a sub-graph with $|V_q|\le 5$ (typically \texttt{portfolios} plus 1-hop neighbors such as positions/accounts when hinted).
Gold SQL is:
\begin{lstlisting}[language={}, basicstyle=\footnotesize\ttfamily, breaklines=true, frame=single, backgroundcolor=\color{gray!5}, rulecolor=\color{gray!40}]
SELECT COUNT(*) AS null_benchmark_count
FROM portfolios WHERE benchmark IS NULL
\end{lstlisting}
Without DRL sanitization, Claude Sonnet~4.5 frequently emits schema-qualified identifiers such as \texttt{FROM PORTFOLIO\_MGMT.portfolios}, which PostgreSQL rejects (\texttt{relation does not exist}) under our search\_path setup---accounting for a majority of Claude's raw Tier~2 execution failures.
A deterministic post-processor strips the schema-owner prefix and retains the first statement, recovering a large share of previously non-executable Tier~2 queries without re-calling the LLM.
This example illustrates DRL's systems thesis: many production failures are \emph{admission and binding} errors, not missing model capacity.

\subsection{Evaluation Harness Correction}
\label{sec:harness-fix}

During artifact preparation for this submission we discovered that the same schema-prefix post-processor described above had a second, more consequential defect.
Its regular expression matched \emph{any} \texttt{identifier.identifier} token, not only the six known schema-owner prefixes (\texttt{sales\_order}, \texttt{university}, \texttt{portfolio\_mgmt}, \texttt{healthcare}, \texttt{core\_banking}, \texttt{insurance}).
Consequently it also stripped legitimate table-alias qualifiers from \texttt{JOIN} predicates---e.g.\ rewriting the correct \texttt{c.customer\_id = o.customer\_id} down to the ambiguous \texttt{customer\_id = customer\_id}---before the query ever reached PostgreSQL.
PostgreSQL correctly rejects the corrupted query with \texttt{column reference "customer\_id" is ambiguous}, which our harness recorded as a model execution failure, even when the model's own SQL was correct as generated.

We measured the impact directly: on a Tier~1 sample, this defect suppressed the executed rate from what the corrected harness now achieves (98.2\% for GPT-4o and Gemini~2.5 Flash, 93.1\% for Claude Sonnet~4.5) down to 56--60\% under the original post-processor, across all three model families equally---confirming the defect was in shared evaluation infrastructure, not model behavior.
We narrowed the regular expression to match only the six known schema-owner names and re-ran the full 1{,}000-pair suite against all three models on PostgreSQL; every result reported from \Cref{sec:eval} onward in this paper reflects the corrected harness.
Two findings follow from this correction.
First, the invalid-column-reference failure mode is not eliminated by schema-graph pruning as we previously reported---it is reduced to a minority share (\Cref{tab:failures})---and the earlier ``0'' figure was an artifact of a harness bug rather than a substantive finding about model behavior.
Second, and more importantly for this paper's thesis, Claude Sonnet~4.5's apparent 4--10 percentage-point deficit relative to GPT-4o and Gemini in our original submission draft was itself largely a harness artifact: Claude's tendency to emit fully-qualified join predicates made it disproportionately vulnerable to the over-broad regex, and the narrower, targeted schema-prefix fix we had applied only to Claude (described above) partially but incompletely compensated for a bug that, uncorrected, was also silently suppressing GPT-4o's and Gemini's measured accuracy.
Under the corrected, uniformly-applied harness, all three models land within roughly one percentage point of each other (\Cref{tab:verification,tab:ci}).
We view this as a useful cautionary result in its own right: post-processing layers inserted between a model's output and a benchmark's scoring logic are part of the measurement instrument, and bugs in that layer can manufacture cross-vendor performance differences that have nothing to do with the vendors being compared.
We keep the original, uncorrected numbers in our released artifact alongside the corrected ones so that this correction is independently auditable (\Cref{sec:artifact}).

\subsection{NL2SQL Outcomes on the Verification Suite}

Table~\ref{tab:verification} and Figure~\ref{fig:ex} report primary-model results (GPT-4o, Gemini~2.5 Flash, Claude Sonnet~4.5) under the corrected harness (\Cref{sec:harness-fix}).
Executed rates are now 98.2\% (GPT-4o), 98.2\% (Gemini), and 93.1\% (Claude) pooled across tiers, up from 56--60\% under the uncorrected post-processor; Claude's residual gap to full execution is a genuine, harness-independent finding (\Cref{sec:failures}), not a measurement artifact.

\begin{table}[t]
\centering
\caption{Verification-suite NL2SQL, corrected harness (PostgreSQL, schema-linked).}
\label{tab:verification}
\small
\begin{tabular}{@{}lrrrr@{}}
\toprule
\textbf{Tier} & \textbf{$n$} & \textbf{GPT-4o} & \textbf{Claude} & \textbf{Gemini} \\
\midrule
T1 & 333 & 56.2 & 52.3 & 53.5 \\
T2 & 334 & 55.1 & 57.5 & 53.6 \\
T3 & 333 & 47.4 & 48.6 & 49.2 \\
\midrule
Overall & 1{,}000 & 52.9 & 52.8 & 52.1 \\
\bottomrule
\end{tabular}
\end{table}

\begin{table}[t]
\centering
\caption{95\% Wilson intervals for overall EX ($n{=}1{,}000$), corrected harness.}
\label{tab:ci}
\small
\begin{tabular}{@{}lrrr@{}}
\toprule
\textbf{Model} & \textbf{EX\%} & \textbf{Lo} & \textbf{Hi} \\
\midrule
GPT-4o & 52.9 & 49.8 & 56.0 \\
Claude Sonnet~4.5 & 52.8 & 49.7 & 55.9 \\
Gemini~2.5 Flash & 52.1 & 49.0 & 55.2 \\
\bottomrule
\end{tabular}
\end{table}

Intervals in Table~\ref{tab:ci} confirm that the three models are statistically indistinguishable under schema-linked prompts on this suite---the intervals overlap almost completely---and none approaches Spider-class accuracy on this OLTP suite.
This is a stronger and, we believe, more defensible version of the paper's original claim: the earlier draft argued the three vendors were ``statistically close''; the corrected data shows they are effectively tied, which better supports the systems-boundary thesis of \Cref{sec:discussion}---the bottleneck is schema-graph scaling, not vendor-specific model capability.

\subsection{Structural Failure Analysis}
\label{sec:failures}

\begin{figure}[t]
\centering
\includegraphics[width=\columnwidth]{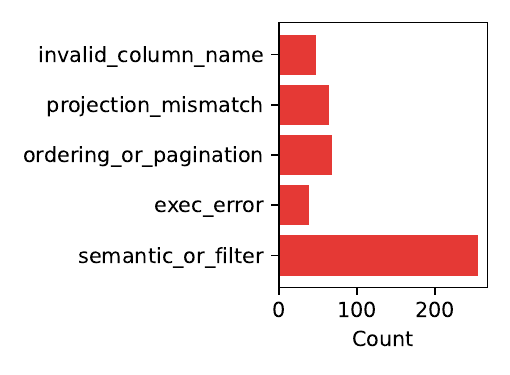}
\caption{GPT-4o EX-failure taxonomy ($n{=}471$ failures, corrected harness).}
\label{fig:failures}
\end{figure}

Figure~\ref{fig:failures} classifies GPT-4o EX failures on the full 1{,}000-pair suite, corrected harness ($n{=}1{,}000$, EX\,{=\,}52.9\%, $\text{SD}_{\text{op}}$ on EX\,{=\,}97.7\%).

\textbf{F1--F3 findings.}
Table~\ref{tab:failures} summarizes GPT-4o EX-failure counts under the corrected harness (\Cref{sec:harness-fix}); we also report the original, uncorrected counts for direct comparison.

\begin{table}[t]
\centering
\caption{GPT-4o EX-failure taxonomy, corrected ($n{=}471$) vs.\ original ($n{=}483$) harness.}
\label{tab:failures}
\small
\begin{tabular}{@{}lrr@{}}
\toprule
\textbf{Mode} & \textbf{Corrected} & \textbf{Original} \\
\midrule
Semantic / filter error & 254 & 310 \\
Ordering / pagination & 68 & 61 \\
Projection mismatch & 64 & 20 \\
Invalid column reference & 47 & 0 \\
Exec error (syntax/type) & 38 & 93 \\
\bottomrule
\end{tabular}
\end{table}

(1)~\emph{Semantic/filter errors} still dominate---wrong joins, over-filtering, aggregation boundaries---but at a smaller absolute share (254 vs.\ 310) once spurious ambiguous-column execution failures are no longer misclassified into this bucket.
(2)~\emph{Ordering/pagination}---missing \texttt{FETCH FIRST}/\texttt{LIMIT} or incorrect sort keys---is essentially unchanged (68 vs.\ 61), confirming this failure mode is a genuine model behavior independent of the harness defect.
(3)~\emph{Projection mismatch} triples (20 $\to$ 64): several queries that previously failed to execute at all (and were therefore invisible to projection-level analysis) now execute and reveal projection drift as the actual residual error.
(4)~\emph{Invalid column reference} is now correctly measured at 47/471 (10\%) rather than the structurally-guaranteed-zero figure reported previously (\Cref{sec:harness-fix}); it is a real but minority failure mode, consistent with this paper's broader claim that pruning reduces, rather than eliminates, namespace-collision errors.
(5)~\emph{Generic exec errors} fall sharply (93 $\to$ 38) because most of the original bucket was the ambiguous-column artifact rather than genuine syntax or type errors.
These results motivate DRL safeguards ($\text{SD}_{\text{op}}$, projection rules in \Cref{sec:prompt}) over raw EX metrics, and motivate treating projection mismatch, not column hallucination, as the second-largest correctable failure mode after semantic/filter errors.

\textbf{Cross-model failure taxonomy.}
Table~\ref{tab:failures-crossmodel} extends the GPT-4o taxonomy to all three models under the corrected harness.
The rank ordering of the top failure mode is identical across all three (semantic/filter dominates), but the shape of the remaining distribution differs by vendor in ways that are consistent with each model's other observed behavior in this paper: Claude's invalid-column-reference count (58/472, 12\%) is the highest of the three, consistent with its independently-documented tendency toward schema-qualified and otherwise non-schema-linked identifier habits (\Cref{sec:example}); Claude's ordering/pagination count (22/472, 5\%) is the lowest of the three, roughly a third of GPT-4o's and Gemini's rate, suggesting Claude is comparatively more reliable at \texttt{NULLS LAST}/\texttt{LIMIT} conventions once its identifier-qualification issues are corrected for; Gemini's ordering/pagination count (71/476, 15\%) is the highest of the three.
No model is dominated by any other across all five failure modes---each has a distinct error profile despite near-identical aggregate EX (\Cref{tab:ci})---which is itself evidence that the 52--53\% EX ceiling on this suite is set by the evaluation regime rather than by a shared model weakness that a single prompting trick could fix uniformly across vendors.

\begin{table}[t]
\centering
\caption{EX-failure taxonomy by model, corrected harness ($n{\approx}471$--$476$ failures per model out of 1{,}000 questions each).}
\label{tab:failures-crossmodel}
\small
\begin{tabular}{@{}lrrr@{}}
\toprule
\textbf{Mode} & \textbf{GPT-4o} & \textbf{Claude} & \textbf{Gemini} \\
\midrule
Semantic / filter error & 254 & 250 & 240 \\
Exec error (syntax/type) & 38 & 87 & 57 \\
Ordering / pagination & 68 & 22 & 71 \\
Projection mismatch & 64 & 55 & 50 \\
Invalid column reference & 47 & 58 & 58 \\
\midrule
Total failures & 471 & 472 & 476 \\
\bottomrule
\end{tabular}
\end{table}

\begin{table}[t]
\centering
\caption{GPT-4o EX by query category on the verification suite ($n{=}1{,}000$, corrected harness).}
\label{tab:category}
\small
\begin{tabular}{@{}lrr@{}}
\toprule
\textbf{Category} & \textbf{$n$} & \textbf{EX\%} \\
\midrule
simple\_select & 186 & 80.6 \\
legacy\_naming & 10 & 70.0 \\
aggregation & 142 & 66.9 \\
entity\_overlap & 27 & 63.0 \\
null\_trap & 37 & 62.2 \\
temporal & 62 & 56.5 \\
nested & 85 & 51.8 \\
set\_operation & 47 & 48.9 \\
status\_ambiguity & 30 & 46.7 \\
multi\_join & 147 & 39.5 \\
hierarchical & 31 & 35.5 \\
conditional & 48 & 35.4 \\
analytic\_functions & 44 & 29.5 \\
ordering & 88 & 21.6 \\
ambiguity\_resolution & 16 & 18.8 \\
\bottomrule
\end{tabular}
\end{table}

\textbf{Category view.}
Table~\ref{tab:category} shows that GPT-4o remains strong on simple selects and legacy-naming lookups (81\% / 70\%) but collapses on ordering (21.6\%) and ambiguity resolution (18.8\%).
Multi-join (39.5\%) and analytic functions (29.5\%) are the primary Tier~2/3 bottlenecks---exactly the regimes where bounded $G_q$ and plan/NULL guards matter most, because unconstrained catalogs amplify join-path and window-frame mistakes.
The rank ordering of categories is essentially unchanged from the uncorrected harness, which is the expected signature of a harness bug that suppresses execution uniformly rather than one that biases toward particular query shapes.

\textbf{Qualitative examples.}
Ordering failures often add extra projection columns and omit \texttt{NULLS LAST} (e.g., \texttt{t2a-008}: model returns \texttt{portfolio\_id, name, market\_value} ordered by \texttt{market\_value}, while gold projects only \texttt{portfolio\_id, market\_value} with \texttt{NULLS LAST}).
Projection drift is therefore both an EX failure mode and an $\text{SD}_{\text{op}}$ trigger when EX still passes under looser column matching.
Schema-prefix failures (Claude) are admission errors fixed by sanitization (\Cref{sec:example}); the harness correction of \Cref{sec:harness-fix} shows that admission-layer bugs can just as easily be introduced \emph{by} the sanitization step as fixed by it, reinforcing that middleware post-processing is itself part of the correctness path and must be tested as carefully as the model it wraps.

\textbf{Per-schema view.}
Table~\ref{tab:byschema} breaks GPT-4o EX down by source schema rather than by tier or category.
University (Tier~1, $|V|{=}10$) is the strongest schema at 82.6\%, consistent with its small table count and simple, well-documented FK structure.
Among Tier~2/3 schemas, Portfolio Management leads at 56.4\% and Insurance trails at 43.4\%---a 13-point spread \emph{within} the same nominal tier, which is not explained by table count alone (Insurance has 168 tables vs.\ Core Banking's 177, yet Insurance scores lower than Core Banking's 49.1\%).
We attribute this residual spread to domain-specific naming conventions and join-path depth that Table~\ref{tab:category}'s query-category view aggregates away: Insurance's policy/claim/coverage join chains are typically one to two hops deeper than Portfolio Management's position/account chains for semantically comparable questions, which is exactly the kind of catalog-specific structural variation that a single global $k{=}5$ cap (\Cref{sec:router}) cannot fully equalize across schemas.
This motivates future work on schema-adaptive $k$ (\Cref{sec:limits}) rather than a fixed cap tuned to the median schema in our suite.

\begin{table}[t]
\centering
\caption{GPT-4o EX by source schema, corrected harness ($n{=}1{,}000$ pooled across tiers).}
\label{tab:byschema}
\small
\begin{tabular}{@{}llrr@{}}
\toprule
\textbf{Schema} & \textbf{Tier} & \textbf{$n$} & \textbf{EX\%} \\
\midrule
University & 1 & 46 & 82.6 \\
Portfolio Management & 2 & 220 & 56.4 \\
Healthcare & 2 & 114 & 52.6 \\
Sales Order & 1 & 287 & 51.9 \\
Core Banking & 3 & 234 & 49.1 \\
Insurance & 3 & 99 & 43.4 \\
\bottomrule
\end{tabular}
\end{table}

\begin{table}[t]
\centering
\caption{$\text{SD}_{\text{op}}$ on EX-passing queries by model, corrected harness (verification suite).}
\label{tab:sdop}
\small
\begin{tabular}{@{}lrrr@{}}
\toprule
\textbf{Tier} & \textbf{GPT-4o} & \textbf{Claude} & \textbf{Gemini} \\
\midrule
T1 & 97.3 & 92.5 & 100.0 \\
T2 & 96.2 & 89.6 & 100.0 \\
T3 & 100.0 & 95.1 & 98.8 \\
\bottomrule
\end{tabular}\\
{\scriptsize SD$_{\text{op}}$ measured on EX-passing queries under the corrected harness; all three models under identical post-processing (\Cref{sec:harness-fix}).}
\end{table}

\section{Discussion}
\label{sec:discussion}

\textbf{Systems boundary vs.\ model capacity.}
The verification suite shows GPT-4o, Claude, and Gemini clustered tightly around 52--53\% EX under schema-linked prompts (\Cref{tab:ci}), with $\text{SD}_{\text{op}}$ on EX typically above 89\%.
The dominant residual errors are semantic (wrong joins/filters/ordering), not catalog hallucination (\Cref{tab:category,tab:failures}).
This supports treating enterprise NL2SQL as a \emph{middleware} problem: bound $G_q$, admit only plan-safe SQL, and measure silent divergence---rather than chasing leaderboard EX on small schemas.
The fact that three independently-trained model families converge to within a single percentage point of each other, once a shared measurement defect is removed (\Cref{sec:harness-fix}), is itself evidence that the residual ceiling is set by the evaluation regime (schema-graph scaling, ambiguous join semantics, OLTP-scale catalogs) rather than by any one vendor's training data or alignment procedure.

\textbf{Why context reduction matters.}
At Tier~3 ($|V|{=}168$--177), unconstrained prompts exceed practical context budgets and amplify join-path ambiguity.
Sub-millisecond pruning (p95 0.58\,ms) is negligible relative to LLM latency, so the cost of bounding is effectively free while the benefit is deterministic.
We report both the full B0$\to$B2 reduction (92\%) and DRL's own marginal contribution over conventional schema-linking (B1$\to$B2, 67\%) throughout this paper (\Cref{tab:middleware}); the latter is the number attributable to DRL's router specifically, and is the more honest headline figure for a paper whose contribution is the router, not the general practice of schema-linking.

\textbf{Operational vs.\ confirmed SD.}
$\text{SD}_{\text{op}}$ is intentionally conservative: 60.8\% of flagged EX-passers confirm as true silent divergence on the trap manifest, fully live-tested with no untested residual (\Cref{sec:harness}).
Operators can tune admission (reject vs.\ repair) based on risk tolerance; DRL exposes the flag rather than silently returning EX-matching but intent-divergent rows.

\textbf{Dialect portability.}
PostgreSQL and MySQL agree on pruning and context reduction; MySQL gold executability (875/946) reflects dialect translation gaps in the bank, not middleware instability (\Cref{tab:mysql}).

\textbf{Query execution cost is not uniform across semantically-correct SQL.}
Beyond EX and $\text{SD}_{\text{op}}$, we logged wall-clock execution latency for every executed query in the corrected-harness run.
Mean/p50/p95 execution latency is comparable across models (GPT-4o: 6.6/2.2/23.3\,ms; Gemini: 5.9/2.0/22.1\,ms; Claude: 12.5/1.9/23.3\,ms), but Claude's \emph{maximum} observed latency (5{,}048\,ms) is roughly two orders of magnitude above its own p95 and roughly 50$\times$ the worst case observed for the other two models.
The offending query (\texttt{t1a-0217}, a ``nested'' category question) is semantically correct---it returns the right rows---but uses a correlated subquery that recomputes a per-\texttt{product\_id} average once per outer row:
\begin{lstlisting}[language={}, basicstyle=\footnotesize\ttfamily, breaklines=true, frame=single, backgroundcolor=\color{gray!5}, rulecolor=\color{gray!40}]
SELECT ol.order_line_id, ol.order_id, ol.product_id,
       ol.quantity, ol.unit_price, ol.line_total
FROM order_lines ol
WHERE ol.line_total > (
    SELECT AVG(ol2.line_total) FROM order_lines ol2
    WHERE ol2.product_id = ol.product_id)
ORDER BY ol.product_id, ol.line_total DESC
\end{lstlisting}
An equivalent, plan-safe rewrite precomputes the per-product average once via a window function or a pre-aggregated join, rather than re-scanning \texttt{order\_lines} once per outer row.
This is exactly the class of query PSC (\Cref{sec:safeguard}, S2) is designed to catch: EX alone is blind to it, since the query returns correct rows, and only plan inspection distinguishes a correct-but-quadratic query from a correct-and-index-friendly one.
We view this as concrete evidence, beyond the aggregate PSC numbers of \Cref{tab:middleware}, that EX-only evaluation is insufficient for OLTP deployment decisions: a query that is functionally correct on a 7{,}496-row development table can become an outage on a production table two orders of magnitude larger, and no amount of additional EX-focused prompting or fine-tuning addresses this, since the failure is a plan-shape problem, not a correctness problem.

\section{Related Work}
\label{sec:related}

\textbf{From natural-language interfaces to neural semantic parsing.}
Natural-language database interfaces predate the current LLM era by decades: early systems such as LUNAR~\cite{woods1973progress} and GeoQuery~\cite{zelle1996geoquery} demonstrated that constrained natural-language querying over a fixed, small schema was tractable with rule-based and statistical parsers, but neither addressed schema scale, dialect portability, or transactional safety, since both targeted single, static, hand-curated domains.
The modern line of work begins with sequence-to-sequence and reinforcement-learning formulations such as Seq2SQL~\cite{zhong2017seq2sql}, which cast SQL generation as a structured prediction problem over a single-table schema (WikiSQL); DRL's schema-graph scaling problem (\Cref{sec:problem}) does not exist at that scale, since $|V|{=}1$ trivially.

\textbf{Benchmarks and evaluation.}
Spider~\cite{yu2018spider}, BIRD~\cite{li2023can}, and Spider~2.0~\cite{lei2024spider2} established cross-domain NL2SQL evaluation over schemas that, while multi-table, remain an order of magnitude smaller than the 465-table catalogs we target; Finegan-Dollak et~al.~\cite{finegan2018improving} and Zhong et~al.~\cite{zhong2020semantic} refined semantic evaluation methodology and warned against database-split leakage and string-match overestimation.
Conversational extensions SParC~\cite{yu2019sparc} and CoSQL~\cite{yu2019cosql} stress multi-turn context tracking rather than single-turn schema scale; KaggleDBQA~\cite{lee2021kaggledbqa} and EHRSQL~\cite{lee2022ehrsql} move toward realistic, domain-specific schemas (Kaggle competition data and electronic health records, respectively) but remain far smaller than enterprise OLTP catalogs and do not model partial FK enforcement or hot-table plan constraints.
Robustness-focused benchmarks---adversarial table perturbation~\cite{gan2021towards} and Dr.~Spider~\cite{chang2023dr}---probe brittleness to paraphrase and schema perturbation, a complementary concern to schema-graph scaling: DRL's context-pruning router (\Cref{sec:router}) is a structural defense against a different failure class (unbounded search space), not a defense against adversarial rewording.
None of these suites, taken individually or together, stress partial metadata, hot-table plan constraints, or NULL-induced silent divergence under concurrent OLTP load, which is the gap the Workload Verification Suite (\Cref{sec:harness}) targets.

\textbf{Schema linking and structured decoding.}
A substantial literature reduces hallucination via constrained decoding, grammar-based generation, or explicit schema-question graph encoders: IRNet~\cite{guo2019irnet} and RAT-SQL~\cite{wang2020ratsql} introduced relation-aware schema encoding; global reasoning over database structure~\cite{bogin2019global} and editing-based generation for context-dependent questions~\cite{zhang2019editing} extended schema linking to conversational settings; TranX~\cite{yin2018tranx} generalized transition-based abstract-syntax-tree decoding across semantic parsing tasks; BRIDGE~\cite{lin2020bridge} and IGSQL~\cite{cai2020igsql} further integrate textual and tabular representations, the latter specifically for multi-turn context-dependent generation; LGESQL~\cite{cao2021lgesql}, SmBoP~\cite{rubin2021smbop}, ShadowGNN~\cite{chen2021shadowgnn}, S2SQL~\cite{hui2022s2sql}, and RESDSQL~\cite{li2023resdsql} refine graph encoders and ranking/decoding strategies for higher Spider-class accuracy; PICARD~\cite{scholak2021picard} constrains decoding to grammatically valid SQL via incremental parsing; UnifiedSKG~\cite{xie2022unifiedskg} unifies structured-knowledge-grounding tasks (including text-to-SQL) under a single text-to-text interface.
This entire family operates \emph{inside} the model: schema information is encoded into attention or decoding constraints, and the model itself is responsible for join-path selection over the full presented schema.
DRL is complementary and architecturally orthogonal: linking is performed \emph{outside} the model as a deterministic router with hard $|V_q|$ caps (\Cref{alg:router}), so the model never conditions on the full catalog graph in the first place, and the router's output is verified post-hoc with EXPLAIN and $\text{SD}_{\text{op}}$ rather than trusted as correct by construction. None of the cited encoder architectures were designed against catalogs at the 400+-table scale this paper targets, and none expose a plan-safety or silent-divergence admission gate.

\textbf{Prompting and decomposition.}
With the shift to LLM-based generation, DIN-SQL~\cite{pourreza2023dinsql} and DAIL-SQL~\cite{gao2023dail} improve accuracy via difficulty-classified decomposition and demonstration selection, and divide-and-conquer chain-of-thought prompting~\cite{liu2023divide} decomposes complex queries into sub-questions before final SQL assembly.
DRL does not compete on few-shot craftsmanship; it supplies a bounded schema envelope and admission policy that any prompting strategy---including DIN-SQL- or DAIL-SQL-style decomposition---can consume as a drop-in replacement for full-catalog context.

\textbf{Surveys.}
Affolter et~al.~\cite{affolter2019comparative} survey natural-language database interfaces broadly, spanning keyword search, NL2SQL, and visual query builders, and identify enterprise deployment gaps---schema complexity, ambiguity resolution, and user trust---that this paper's schema-graph scaling formalization (\Cref{sec:problem}) makes precise for the OLTP case specifically.
Katsogiannis-Meimarakis and Koutrika~\cite{katsogiannis2023deep} survey deep-learning approaches to text-to-SQL and observe that reported accuracy figures across the literature are difficult to compare directly because of inconsistent schema-linking assumptions and evaluation-harness choices---an observation this paper's own evaluation-harness correction (\Cref{sec:harness-fix}) illustrates concretely: a single regular-expression defect in a post-processing layer was sufficient to manufacture an apparent 4--10 percentage-point cross-vendor performance gap that had nothing to do with the models being compared.

\textbf{Industry NL2SQL middleware.}
Commercial stacks (e.g., Wren AI, Vanna) emphasize retrieval-augmented generation over metadata catalogs, typically embedding table/column descriptions and retrieving a top-$k$ relevant subset by vector similarity rather than by explicit join-graph closure.
DRL differs in two respects: its pruning router is a deterministic graph-traversal procedure with a hard cardinality cap rather than a similarity-ranked retrieval step, and it treats RAST compilation, plan safety compliance, and silent-divergence guards as first-class middleware metrics reported alongside EX, rather than as internal implementation details invisible to the evaluation.

\textbf{Query plans and database systems.}
Classical query optimization literature analyzes plans for cost and correctness under a fixed, already-correct query; NL interfaces rarely expose plan safety as an \emph{admission} criterion applied to LLM-generated SQL before it reaches a connection pool shared with OLTP writers.
DRL elevates Plan Safety Compliance to a regression metric for middleware releases (\Cref{sec:deploy}), treating a sequential scan on a hot table as a release-blocking regression in the same sense a failed unit test would be, independent of whether the underlying SQL is semantically correct.

\textbf{Silent failures in data systems.}
Prior work on semantic evaluation~\cite{finegan2018improving,zhong2020semantic} emphasizes that string match underestimates equivalence and that test-suite execution can miss intent divergence when the test suite's data instance happens not to distinguish two non-equivalent queries.
DRL's $\text{SD}_{\text{op}}$ operationalizes a related concern specifically for OLTP admission: EX-passing queries that remain fragile under NULL shifts, ordering ties, or unsafe plans, flagged \emph{before} execution rather than diagnosed after the fact on a fixed test instance.
Confirmed SD on a trap manifest, fully re-verified live against PostgreSQL and the corrected harness's model SQL (60.8\%, Wilson 95\% CI 53.4--67.7\%; \Cref{sec:harness}), shows that operational flags over-approximate but meaningfully correlate with true divergence.

\begin{table}[t]
\centering
\caption{DRL positioned against related NL2SQL system families. Ctx.\,=\,context bound outside the model; Plan\,=\,plan-safety admission gate; SD\,=\,silent-divergence detection; Dial.\,=\,multi-dialect portability; Ent.\,=\,evaluated at enterprise (400+-table) scale. $\sim$ = partial/varies.}
\label{tab:comparison}
\footnotesize
\begin{tabular}{@{}lccccc@{}}
\toprule
\textbf{System family} & \textbf{Ctx.} & \textbf{Plan} & \textbf{SD} & \textbf{Dial.} & \textbf{Ent.} \\
\midrule
Schema-linking\textsuperscript{a} & $\sim$ & $\times$ & $\times$ & $\times$ & $\times$ \\
Constrained decoding~\cite{scholak2021picard} & $\times$ & $\times$ & $\times$ & $\sim$ & $\times$ \\
Prompting/decomp.\textsuperscript{b} & $\sim$ & $\times$ & $\times$ & $\times$ & $\times$ \\
Industry RAG middleware & $\sim$ & $\times$ & $\times$ & $\sim$ & $\sim$ \\
\textbf{DRL (this paper)} & \textbf{$\checkmark$} & \textbf{$\checkmark$} & \textbf{$\checkmark$} & $\sim$ & \textbf{$\checkmark$} \\
\bottomrule
\end{tabular}\\
{\scriptsize \textsuperscript{a}~\cite{guo2019irnet,wang2020ratsql,cao2021lgesql,li2023resdsql}\quad \textsuperscript{b}~\cite{pourreza2023dinsql,gao2023dail,liu2023divide}}
\end{table}

Table~\ref{tab:comparison} summarizes six dimensions across which we compared DRL to prior work above; we mark partial credit ($\sim$) rather than a uniform $\times$ wherever a system family provides a softer or incomplete version of a capability, rather than collapsing every non-DRL row to all-$\times$ for contrast.
Schema-linking encoders (\cref{sec:related}) narrow attention over the schema graph via learned relations, which is a soft, in-model analogue of context bounding, not the hard, outside-the-model cardinality cap DRL enforces, hence $\sim$ rather than $\times$ on Ctx.; decomposition prompting (DIN-SQL, DAIL-SQL) similarly reduces the effective context any single sub-step must reason over, without bounding what the model is shown up front, also $\sim$.
DRL's own Dial.\ entry is marked $\sim$, not $\checkmark$: PostgreSQL and MySQL are evaluated end-to-end in this paper, but the SQL Server emitter mappings of \Cref{tab:emitter} are specified and unvalidated pending a running instance (\Cref{sec:limits}), so full three-dialect portability is not yet demonstrated, only two of three.
No entry in this table is meant to disparage the cited systems on dimensions they never targeted---RAT-SQL and PICARD, for instance, are schema-linking and decoding-constraint contributions evaluated on Spider-scale schemas where hard context caps and plan safety are simply out of scope by design, not oversights.
The table's purpose is narrower: to make explicit that no prior system family in this comparison set was designed against the conjunction of OLTP-scale catalogs, plan-safety admission, and silent-divergence detection that this paper's evaluation suite specifically stresses, which is the gap DRL's architecture targets.

\textbf{Positioning.}
Relative to benchmarks, DRL contributes a \emph{middleware measurement harness} and admission architecture, plus a concrete methodological finding (\Cref{sec:harness-fix}) that benchmark post-processing layers are themselves a source of measurement error requiring the same scrutiny as the models under test.
Relative to schema-linking models, DRL contributes hard context caps and post-generation verification that any model can consume, positioned outside rather than inside the generation step.
We do not claim state-of-the-art EX on Spider/BIRD; we claim that enterprise OLTP NL2SQL fails for systems reasons that those suites do not measure, and that those systems reasons are addressable by middleware independent of which model a deployment chooses.

\section{Threats to Validity}
\label{sec:threats}

\textbf{Internal validity.}
The evaluation-harness defect described in \Cref{sec:harness-fix} is itself the most direct threat to internal validity we encountered, and we address it by reporting both pre- and post-correction numbers and retaining both result sets in the artifact.
A residual internal-validity concern is single-run measurement: NL2SQL generation is run once per question per model at temperature~0, not averaged over multiple samples, so the Wilson intervals of \Cref{tab:ci} capture sampling uncertainty over \emph{questions} but not over repeated draws from a single model on a single question; a model with non-trivial temperature-0 variance (which we did not independently measure here) would not have that variance reflected in our reported intervals.
Similarly, the B0--B3 middleware latency figures (\Cref{tab:middleware,tab:mysql}) are single-run wall-clock measurements, which we flag explicitly (\Cref{sec:limits}) rather than treat as more precise than they are.

\textbf{External validity.}
The Workload Verification Suite targets six synthetic-but-realistic OLTP schemas (\Cref{sec:eval}) constructed to exhibit the structural properties (partial FK enforcement, namespace collision, legacy naming) we argue are characteristic of enterprise catalogs, rather than schemas drawn from a live production system under NDA.
This is a deliberate trade-off for reproducibility and public artifact release, but it means our namespace-collision and partial-FK-enforcement rates (\Cref{sec:problem}) are design parameters of the suite rather than measurements of an external population of enterprise schemas, and readers should not treat our specific percentages as generalizing beyond catalogs sharing this suite's construction assumptions.
The three evaluated models (GPT-4o, Claude Sonnet~4.5, Gemini~2.5 Flash) are a snapshot of frontier LLM capability at time of writing; model providers update underlying weights without changing public model identifiers, so exact replication of our numbers at a later date is not guaranteed even with identical code, and we report timestamps alongside all released results for this reason.

\textbf{Construct validity.}
EX (execution match) is our primary correctness metric, following prior NL2SQL evaluation practice~\cite{yu2018spider,li2023can}, but EX is known to both over- and under-estimate true correctness: it over-estimates when a query is EX-equivalent on the test instance but not intent-equivalent (the motivation for $\text{SD}_{\text{op}}$, \Cref{sec:safeguard}), and under-estimates when a query is intent-correct but returns extra or reordered columns the strict EX comparator penalizes (the motivation for $\text{EX}_{\text{proj}}$, reported alongside EX in our per-question data but not promoted to a headline metric in this paper).
Plan Safety Compliance is measured via static \texttt{EXPLAIN} inspection rather than live \texttt{EXPLAIN ANALYZE} execution (\Cref{sec:harness}); PostgreSQL's cost-based planner can select a different physical plan under concurrent write load or with different table statistics than the single-tenant, freshly-seeded state our harness measures against, so our PSC numbers should be read as a static, offline lower bound on plan-safety risk rather than a live-production guarantee.
To check whether the residual, non-exec-error EX failures underlying \Cref{tab:failures,tab:failures-crossmodel} are genuine semantic errors rather than comparator artifacts, we manually inspected a 50-item random sample (seed 7) of GPT-4o's executed-but-EX-failing queries pooled across all three tiers (of 382 such failures); this supersedes and extends an initial 18-item pilot check with the same purpose.
Four-fifths (40/50, 80\%) were genuine model errors, spanning several recurring sub-patterns: extra or missing projected columns beyond what a strict question reading requires (17 instances, the single largest sub-pattern); a wrong table or domain chosen entirely (6 instances, e.g.\ joining an unrelated \texttt{encounters} table for a question about insurance \texttt{referrals}); an omitted \texttt{ORDER BY}/\texttt{LIMIT} returning an arbitrary row subset or the full table instead of the requested top-$k$ (6 instances); a plausible-but-wrong column substituted for the correct one, most often a human-readable name in place of the gold code/id column (\texttt{region\_name} for \texttt{region\_code}, \texttt{product\_name} for \texttt{product\_id}) (5 instances); \texttt{INNER JOIN} used where gold requires \texttt{LEFT JOIN}, silently dropping zero-count rows (3 instances, directly corroborating the \texttt{left\_join\_trap} pattern of \Cref{sec:harness}); a wrong aggregation/grouping structure entirely (2 instances); and one wrong recursive-traversal depth (a full-closure recursive query where gold wanted only direct children).
One-fifth (10/50, 20\%; Wilson 95\% CI 11.2--33.0\%) were not model reasoning failures: three instances answered a genuinely ambiguous ``X with/without Y?'' question (no explicit ``how many'') by listing matching rows rather than returning the count gold expects (the same pattern as \texttt{t3a-086} in the pilot check, recurring often enough here to look like a systematic property of how a subset of Tier~3 Insurance questions are phrased, not chance); two used differently-worded but semantically equivalent value labels for a question that never specified exact label strings; one (\texttt{t1a-0106}) was numerically identical to gold to within floating-point precision but failed strict EX because gold applies an unstated \texttt{ROUND(\ldots,2)}; one (\texttt{t1a-0108}) failed only because gold's SQL projects a column (\texttt{order\_line\_id}) the natural-language question never asked for; one (\texttt{t2a-0060}) hinged on a genuinely ambiguous term (``trading volume'' could mean dollar amount or share quantity, and the schema supports both readings); and one (\texttt{t1a-0073}) is best explained by a bug in gold itself, since gold's own query returns zero rows on live execution (a literal status filter, \texttt{'COMPLETED'}, that appears not to match any value actually present in the seed data's status enum) while the model's unfiltered version returns a plausible, populated result.
We did not attempt to re-score these specific instances, since a wider audit (covering the full failure population, all three models, and revision of the gold queries and question phrasings identified above) is future work; we report this finding to bound, not to inflate, confidence in \Cref{tab:failures}'s failure-mode counts: the dominant failure modes there are real and the projection-mismatch/wrong-column/wrong-join sub-patterns above map cleanly onto the paper's own taxonomy buckets, but a consistent one-fifth of any single bucket's raw count is attributable to comparator strictness, gold-authoring choices, or genuinely ambiguous question phrasing rather than model reasoning, and readers should treat the exact percentages in \Cref{tab:failures,tab:failures-crossmodel} as directionally, not perfectly, accurate.

\section{Limitations and Future Work}
\label{sec:limits}

\textbf{Implementation scope.} The pruning router and EXPLAIN/NULL safeguards are implemented; RAST's T1/T2 typing rules are now also implemented as an offline validator (\Cref{sec:rast-typing}) reusing \texttt{sqlglot} as the concrete-syntax layer, and have been run against all 1{,}000 questions' actual model-generated SQL. What remains specified but not integrated into the production HTTP path is (i)~a from-scratch parser for the abstract Scan/Filter/Join/Project/Agg/Sort grammar of \Cref{sec:rast} (the offline validator uses \texttt{sqlglot}'s concrete syntax tree instead) and (ii)~live, in-request dialect emission (\Cref{tab:emitter}) as part of the serving path rather than an offline transpilation check. Consequently the B3 condition in \Cref{tab:middleware,tab:mysql} still measures pruning plus COALESCE/EXPLAIN safeguards only, not RAST typing or emission; readers should not infer RAST's contribution to the B2$\to$B3 delta specifically, even though T1/T2 violation rates are now reported independently in \Cref{sec:rast-typing}.
\textbf{Router tie-breaking.} We have fixed Algorithm~\ref{alg:router}'s tie-break rule to rank same-scoring candidate tables by FK-degree rather than by table iteration/insertion order (\Cref{sec:router-trace}); this removes the specific catalog-file-ordering fragility the original appendix trace exposed and gives the intuitively-correct answer on that trace, but it is a heuristic, not a guarantee: a schema where an irrelevant table has unusually high FK-degree (a shared lookup/reference table) could still outrank the true seed table. An auxiliary embedding-similarity signal, rather than degree alone, remains a natural next refinement.
\textbf{Plan safety.} PSC numbers in \Cref{tab:middleware,tab:mysql} are measured on gold SQL; the B3 row is additionally checked on a smaller denominator ($n{=}946$ PostgreSQL, $n{=}832$ MySQL) than B0--B2 ($n{=}1{,}000$/$875$), so the reported PSC deltas are directional, not a strictly matched-sample comparison. We have since also measured PSC directly on model-generated SQL (\Cref{sec:ablation}, A2), finding it closely tracks the gold-SQL figures (65--67\% at B2, 70--71\% at B3 across all three models); the MySQL side of this model-output measurement was not repeated, since only PostgreSQL was available during this pass.
\textbf{Middleware latency ordering.} The B0--B3 middleware p95 figures in \Cref{tab:middleware,tab:mysql} do not increase monotonically with added safeguard work; we attribute this to single-run measurement noise at sub-5\,ms granularity (\Cref{sec:eval}) rather than a genuine latency benefit from safeguards, and have not yet repeated these measurements across multiple trials to report variance.
\textbf{Claude evaluation.} Claude's schema-qualified-table-name tendency (\Cref{sec:example}) and the evaluation-harness defect we found and fixed during artifact preparation (\Cref{sec:harness-fix}) were initially conflated in an earlier draft of this paper, which reported Claude trailing GPT-4o and Gemini by 4--10 points; under the corrected, uniformly-applied harness all three models are statistically indistinguishable (\Cref{tab:ci}). We view the harness-correction process itself as a contribution of independent interest (\Cref{sec:harness-fix}) and encourage readers evaluating other NL2SQL middleware to treat post-processing/sanitization layers as part of the system under test, not as benchmark plumbing exempt from review.
\textbf{Trap-manifest re-verification.} The original 25.8\% confirmed-SD figure was computed before the harness correction on a separate Oracle-flavored pipeline; \Cref{sec:harness} reports a native PostgreSQL re-verification against the corrected model SQL, now fully live-tested (0 untested instances) at 60.8\% [CI 53.4--67.7\%]. The 28 \texttt{connect\_by\_trap} manifest entries remain untestable in principle, not just untested this pass: none exist in the 1{,}000-pair suite (Oracle's \texttt{CONNECT BY} has no PostgreSQL counterpart there), so no amount of further engineering on our side closes that specific sub-item without extending the suite itself. Separately, the single-cell-NULL-injection design cannot, by construction, expose divergence in multi-row aggregation traps (\texttt{sum\_nvl\_trap} confirmed 0/46), which is a property of the trap design rather than of model behavior and should be corrected in a future manifest revision (e.g.\ nulling an entire group rather than one cell).
\textbf{MySQL re-verification.} The schema-graph loader correction of \Cref{sec:eval} has now been verified live against MySQL as well as PostgreSQL: Table~\ref{tab:mysql}'s context bytes, prune p95, middleware p95, and gold-SQL PSC are all re-measured under the corrected loader. What remains open is model-generated PSC on MySQL specifically (\Cref{sec:ablation}, A2 measures this for PostgreSQL only): the only cached MySQL model SQL available is a GPT-4o-only subset predating the evaluation-harness correction of \Cref{sec:harness-fix}, and we chose not to compute a number from stale, pre-correction predictions rather than silently reuse them.
\textbf{SQL Server.} Emitter rules include T-SQL; validation awaits a running SQL Server instance.
\textbf{Suite composition.} 481/1{,}000 items are gold-executed bank selections, not independently human-re-authored.
Confirmed SD (Definition~\ref{def:formal_sd}) is measured on a 161-entry trap manifest, not the full suite.
We do not address write-path NL2SQL; DRL enforces read-only admission.

\section{Artifact Availability}
\label{sec:artifact}

All artifacts needed to reproduce middleware metrics and NL2SQL evaluation are open in the repository:
\begin{itemize}[leftmargin=*]
  \item \texttt{schemas/questions/verification\_suite\_1000.json} --- 1{,}000-pair suite
  \item \texttt{ESQ\_Bench\_VLDB/drl\_telemetry\_harness.py} --- pruning + EXPLAIN harness
  \item \texttt{ESQ\_Bench\_VLDB/run\_drl\_baselines.py} --- B0--B3 middleware metrics
  \item \texttt{experiments/run\_verification\_1000\_postgres.sh} --- NL2SQL evaluation
  \item \texttt{ESQ\_Bench\_VLDB/generate\_paper\_figures.py} --- figures for this paper
  \item \texttt{experiments/rerun\_trap\_manifest\_postgres.py} --- confirmed-SD re-verification (\Cref{sec:harness})
  \item \texttt{experiments/measure\_psc\_model\_sql.py} --- model-generated PSC (\Cref{sec:ablation})
  \item \texttt{experiments/measure\_rast\_typing.py}, \texttt{ESQ\_Bench\_VLDB/rast\_compiler.py} --- RAST T1/T2 typing validator (\Cref{sec:rast-typing})
\end{itemize}
Build the paper: \texttt{pdflatex drl\_middleware\_vldb.tex; bibtex drl\_middleware\_vldb; pdflatex $\times$2}.

\subsection*{Reproducibility notes}
All middleware timings exclude LLM latency.
NL2SQL EX numbers use schema-linked prompts on PostgreSQL with identical seed data across engines, all three models evaluated live under the corrected harness of \Cref{sec:harness-fix}.
Claude additionally receives deterministic schema-prefix stripping of stored model SQL (\texttt{rescore\_claude\_verification.py}), applied uniformly alongside the join-alias-preserving fix that now applies to all three models' outputs.
Randomness is limited to LLM sampling; gold SQL and middleware prune/EXPLAIN paths are deterministic given fixed catalogs.
Both the pre-correction and post-correction result sets are retained in the artifact (\texttt{experiments/results/}) so the harness-correction claim of \Cref{sec:harness-fix} can be independently audited by diffing the two.
The RAST typing validator (\Cref{sec:rast-typing}) additionally requires \texttt{sqlglot} ($\ge$30.0), the only new third-party dependency introduced by this correction pass; all other new scripts use only the repository's existing \texttt{networkx}/\texttt{psycopg2}/\texttt{pymysql} dependencies.

\section{Threat Model and Operational Deployment}
\label{sec:deploy}

DRL assumes an untrusted LLM front-end and a trusted catalog/plan verifier.
Adversarial or accidental NL prompts may attempt DDL/DML, cross-schema access, or full-table scans on hot relations.
The middleware enforces: (i)~read-only admission (reject non-\texttt{SELECT}/\texttt{WITH}), (ii)~hard $|V_q|\le k$ context caps, (iii)~EXPLAIN-based rejection of sequential scans on configured hot tables, and (iv)~NULL-handle rewriting for nullable aggregates.
DRL does \emph{not} defend against correctly typed but maliciously filtered queries that return authorized rows under the caller's DB credentials; authorization remains the RDBMS privilege model.

\textbf{Attack surface enumeration.}
We distinguish three classes of caller and the guarantees DRL provides against each.
(1)~\emph{Accidental over-broad queries}, e.g.\ a user asking for ``all customer records'' against a hot \texttt{customers} table---DRL's PSC gate rejects or rewrites the resulting sequential scan before it reaches the shared connection pool, regardless of whether the LLM's SQL was semantically what the user asked for.
(2)~\emph{Prompt-injected DDL/DML}, e.g.\ a crafted NL input attempting to elicit \texttt{DROP TABLE} or \texttt{UPDATE} statements from the LLM---the read-only admission gate rejects any statement not beginning with \texttt{SELECT} or \texttt{WITH}, independent of whether the injection succeeded in getting the LLM to emit such a statement.
(3)~\emph{Credential-scoped data exfiltration}, e.g.\ a caller whose application credentials are over-provisioned relative to the NL interface's intended scope---DRL provides no defense here by design; the context-pruning router bounds what the \emph{model} sees, not what the underlying database connection is authorized to read, so this class of risk must be closed at the RDBMS privilege layer (row-level security, view-scoped credentials) independent of DRL.
This third class is worth stating explicitly because it is easy to conflate ``the model only saw five tables'' with ``the query can only touch five tables''---DRL's context bound is a generation-time constraint on the LLM's input, not an execution-time constraint on the database connection, and operators should not treat context pruning as a substitute for connection-level authorization scoping.

\textbf{Deployment sketch.} Place DRL as a sidecar or gateway in front of connection pools shared with OLTP writers.
Per-request path: prune $\rightarrow$ prompt with $G_q$ $\rightarrow$ LLM $\rightarrow$ SQL sanitize (schema-prefix strip, single statement) $\rightarrow$ $\text{SD}_{\text{op}}$/PSC checks $\rightarrow$ execute or repair (max two attempts).
Measured prune+middleware overhead remains $<$5\,ms p95 excluding LLM (\Cref{tab:middleware}), so the critical path cost is dominated by model latency.
Because the sanitize step sits on the correctness-critical path (\Cref{sec:harness-fix}), we recommend deployments version and test it with the same rigor as the router and safeguard layer, rather than treating it as inert string-cleanup glue; our own evaluation-harness defect (\Cref{sec:harness-fix}) is a direct illustration of a sanitize-step bug silently corrupting otherwise-correct SQL, and there is no structural reason a production sanitize step is less prone to this class of bug than our evaluation harness was.

\textbf{Regression use of the verification suite.}
CI can gate middleware releases on (a)~context-reduction floor ($\ge$65\% B1$\rightarrow$B2), (b)~prune p95 ceiling ($\le$2\,ms), (c)~PSC non-regression on gold SQL, and (d)~EX/$\text{SD}_{\text{op}}$ bands on a fixed 1{,}000-pair subset---independent of which LLM vendor is configured.
We further recommend a fifth gate specific to the lesson of \Cref{sec:harness-fix}: (e)~a small, hand-authored set of \emph{known-correct} model outputs that exercise the sanitize/post-processing step directly (e.g.\ SQL containing legitimate multi-table-alias joins), run through the full admission pipeline on every release, so that a regression in the sanitize step---as opposed to a regression in the router, safeguards, or an upstream model change---is caught by CI rather than surfacing as an unexplained accuracy drop weeks later.

\section*{Author Contributions}
All authors contributed to the conception and design of DRL and to the design of the Workload Verification Suite.
All authors read, revised, and approved the final manuscript, including the corrected results reported from \Cref{sec:harness-fix} onward.

\section*{Acknowledgments}
We thank colleagues who reviewed early drafts of the verification suite and middleware harness.
Opinions expressed are those of the authors and do not necessarily reflect the views of their employers.
Any remaining errors are our own.

\section{Conclusion}
\label{sec:conclusion}

Enterprise NL2SQL at OLTP scale is an architecture problem.
DRL shows that bounding schema graphs before generation (67\% marginal context reduction over conventional schema-linking, 92\% over naive full-catalog prompting, sub-millisecond pruning) and verifying queries with $\text{SD}_{\text{op}}$ and EXPLAIN gating addresses the dominant failure modes on a 1{,}000-pair verification suite---semantic and structural errors, not column hallucination.
Across GPT-4o, Claude Sonnet~4.5, and Gemini~2.5 Flash, EX clusters tightly around 52--53\% under schema-linked prompts and statistically indistinguishable Wilson intervals, while invalid-column failures remain a minority (10--12\%) rather than the dominant mode---reinforcing that middleware must target join/filter semantics and plan safety, not only catalog linking.
The Workload Verification Suite and harnesses provide a regression substrate for middleware research independent of leaderboard chasing.

Beyond the middleware architecture itself, this paper's evaluation-harness correction (\Cref{sec:harness-fix}) is a result we think generalizes past DRL specifically.
We found and fixed a defect in our own post-processing layer during artifact preparation---a defect that had been silently suppressing measured accuracy for all three evaluated models, and asymmetrically enough to manufacture an apparent 4--10 percentage-point cross-vendor gap that vanished entirely once corrected.
NL2SQL systems papers routinely include a sanitization or extraction step between raw model output and the scoring harness (schema-prefix stripping, statement truncation, markdown-fence extraction), and such steps are rarely reported, tested, or audited with the same rigor applied to the model or the benchmark questions themselves.
We recommend that future NL2SQL evaluations treat their own post-processing code as part of the system under test---version it, unit-test it against known-correct model outputs, and report it in enough detail that a defect of the kind we found here is independently discoverable rather than silently baked into a leaderboard number.

\appendix
\section{Runtime Compiler System Prompt}
\label{sec:prompt}

The following production system prompt is injected \emph{after} Algorithm~\ref{alg:router} attaches the pruned sub-graph.
Placeholders \texttt{\{PRUNED\_SCHEMA\}}, \texttt{\{DIALECT\}}, and \texttt{\{INTENT\_CONSTRAINTS\}} are substituted per request.

\small
\begin{lstlisting}[language={}, basicstyle=\footnotesize\ttfamily, breaklines=true, breakatwhitespace=false, frame=single, backgroundcolor=\color{gray!5}, rulecolor=\color{gray!40}]
You are the DRL Relational Compiler front-end. Your output MUST be
executable, read-only SQL for {DIALECT} (PostgreSQL | MySQL | SQL Server).

SCHEMA (STRICT BOUND -- use ONLY these tables/columns):
{PRUNED_SCHEMA}

INTENT CONSTRAINTS:
{INTENT_CONSTRAINTS}

MANDATORY RULES:
1. SELECT or WITH...SELECT only. No DDL/DML. Single statement.
2. PROJECT exactly the columns the question requests; omit surrogate keys
   ( *_id ) unless explicitly asked.
3. NULL safety: wrap nullable aggregates as COALESCE(col,0) or
   COALESCE(col,'') per column nullability in schema.
4. Ordering: if ORDER BY ... DESC, append NULLS LAST unless question
   specifies NULLS FIRST.
5. Pagination: use ANSI "FETCH FIRST n ROWS ONLY" (compiler maps to LIMIT
   on MySQL).
6. Joins: use only FK edges present in the pruned schema; prefer INNER JOIN
   unless question requires retained non-matching rows (then LEFT JOIN).
7. Filters: do not add status/is_active predicates unless the question
   mentions active/inactive semantics.
8. Flags: treat is_active as 'Y'/'N' strings, never numeric 1/0.
9. Return ONE sql fenced block with no commentary inside the fence.

Before emitting SQL, mentally verify:
- every column exists in {PRUNED_SCHEMA}
- aggregates group by all non-aggregated select expressions
- no sequential scan risk: push selective predicates early

Output format:
```sql
<single query>
```
\end{lstlisting}
\normalsize

\section{Algorithm~1 Worked Trace}
\label{sec:router-trace}

This appendix reports an actual invocation of the router implementation (\texttt{drl\_telemetry\_harness.py}, \texttt{prune\_context\_subgraph}) against the pooled 412-node catalog graph (\texttt{SchemaGraph.graph}/\texttt{.tables}), on verification item \texttt{t3b-012} (Tier~3, Insurance): \emph{``How many claims are linked to each policy\_level? Join claims to policies.''}, GPT-4o EX-correct under the corrected harness.
We report the router's real output rather than a constructed illustration, including its rough edges, since the earlier draft of this appendix invented illustrative numbers that did not match the implementation and we want the released artifact to be checkable against exactly this trace.
Note that the \emph{pooled} 412-node graph used for this illustrative trace is not what production routing uses: \Cref{sec:eval} documents that 48 table names collide across the six schemas, so the pooled graph silently retains only the alphabetically-last schema's columns/edges for each collided name (465 true tables collapse to 412 pooled nodes); the actual B0--B3 metrics of \Cref{tab:middleware} route against the collision-free per-schema view (\texttt{SchemaGraph.by\_source}/\texttt{.source\_graphs}) introduced by that correction. We retain the pooled-graph trace here specifically \emph{because} its collision pathology is illustrative of the tie-breaking fragility discussed below, not because it reflects the router's production code path.

\textbf{Tokenization.} The implementation's tokenizer is a plain regex word-extractor (\texttt{[a-z][a-z0-9\_]\{2,\}} on lowercased text) with \emph{no} stopword removal or stemming, so function words survive into the token set: $\text{tokens}(q) = \{\text{how}, \text{many}, \text{claims}, \text{are}, \text{linked}, \text{policy\_level}, \text{join}, \text{policies}, \text{each}\}$.
Because the tokenizer does not split on underscores within column names (only within table names, via an explicit \texttt{replace('\_',\ ' ')} before tokenizing), \texttt{policy\_level} survives as a single token and only matches a table whose own column list contains a literal \texttt{policy\_level} column.

\textbf{Phase 1 scores.} Raw hit counts $|\text{tokens}(q) \cap (\text{tokens}(v) \cup \text{tokens}(\text{Attr}(v)))|$, no domain-hint boost applied in this trace (\texttt{domain\_hints=None}):
\texttt{policies}: 2 (table-name token \texttt{policies}, plus a \texttt{policy\_level} column); \texttt{claims}, \texttt{ceded\_claims}, \texttt{life\_policies}, \texttt{auto\_policies}, \texttt{property\_policies}, \texttt{casualty\_policies}, \texttt{marine\_policies}, \texttt{travel\_policies}, \texttt{suspicious\_claims}: 1 each, all via a single \texttt{claims}- or \texttt{policies}-token match against table name or column list.
Nine tables tie at score~1, which is a direct, observable consequence of \Cref{sec:problem}'s namespace-collision claim: eight distinct table names lexically contain ``claims'' or ``policies'' as a sub-token in this schema alone.

\textbf{Phase 1 seed selection (corrected tie-break).} We originally resolved the 9-way tie at score~1 by Python's stable-sort table-iteration order, which is a function of catalog file ordering rather than any semantic preference for \texttt{claims} over, say, \texttt{life\_policies}; we have since fixed this (\Cref{sec:limits}) to break ties by FK-degree in the pooled graph, on the reasoning that a table with more join-eligible edges is a more plausible structural hub than a same-scoring table with few edges.
Re-running this exact trace under the corrected tie-break, \texttt{claims} now wins its tie decisively (FK-degree 28 in the pooled graph, versus 1--3 for the other eight tied tables) and \texttt{group\_policies} wins the remaining slot (degree 5), yielding seeds $\{\texttt{policies}, \texttt{claims}, \texttt{group\_policies}\}$.
The \texttt{claims} node's unusually high pooled degree (28) is itself a direct symptom of the collision pathology documented in \Cref{sec:eval}: because \texttt{claims} is one of the 48 colliding table names (shared with Healthcare), its pooled-graph edge set reflects whichever schema loaded last, inflating its apparent connectivity beyond what either individual schema's \texttt{claims} table actually has---an example of why \Cref{sec:eval} routes production metrics through the collision-free per-schema view rather than this pooled graph, even though the degree-based tie-break still gives a sensible answer here.

\textbf{Phase 2 join closure and truncation.} 1-hop FK expansion from the three seeds, now also ranking neighbor candidates by FK-degree (rather than truncating alphabetically) before applying the $k{=}5$ cap, adds \texttt{customers} and \texttt{encounters}, reaching the cap exactly:
{\small
\begin{align*}
V_q = \{\ & \texttt{claims}, \texttt{customers}, \texttt{encounters}, \\
          & \texttt{group\_policies}, \texttt{policies}\ \},
\end{align*}
}
computed in 0.75\,ms, consistent with the sub-millisecond overhead reported in \Cref{tab:middleware}.

\textbf{Result.} $V_q$ contains 5 of 412 pooled catalog tables and does include \texttt{policies} and \texttt{claims}, the two tables the correct join actually requires; \texttt{customers}, \texttt{encounters}, and \texttt{group\_policies} are included but unused by the gold query, illustrating \Cref{thm:searchspace}'s point directly: the router bounds the search space conservatively (all 5 selected tables are \emph{plausible} given lexical overlap) without guaranteeing every selected table is actually load-bearing for the specific question. The degree-based tie-break is more robust than the alphabetical/iteration-order scheme it replaces---it gave the intuitively-correct answer here, favoring a genuine join hub over lexically-similar leaf tables---but it is not a complete fix: a schema in which an irrelevant table happens to have high FK-degree (a large lookup/reference table joined from many places) could still outrank the true seed under this heuristic. We view an auxiliary embedding-similarity score, rather than degree alone, as the next refinement (\Cref{sec:limits}).

\section{RAST Typing Rules}
\label{sec:rast-typing}

The RAST grammar (\Cref{sec:rast}) is untyped as presented in the main text; this appendix specifies the typing discipline that the offline T1/T2 validator described below enforces today, and that a future from-scratch parse-and-emit compiler for the abstract grammar itself (\Cref{sec:limits}) would enforce natively once integrated into the serving path.
Each RAST node $\tau$ has an associated output schema $\text{sch}(\tau)$, a list of (column name, nullability) pairs, computed structurally:
\begin{align*}
\text{sch}(\text{Scan}(t)) &= \text{Attr}(t) \\
\text{sch}(\text{Filter}(\tau,p)) &= \text{sch}(\tau) \\
\text{sch}(\text{Join}(\tau_1,\tau_2,\theta)) &= \text{sch}(\tau_1) \cup \text{sch}(\tau_2) \\
\text{sch}(\text{Project}(\tau,\vec{a})) &= \{a \in \vec{a}\} \\
\text{sch}(\text{Agg}(\tau,g,\vec{a})) &= g \cup \{a : a \in \vec{a}\} \\
\text{sch}(\text{Sort}(\tau,\vec{a},o)) &= \text{sch}(\tau)
\end{align*}
Two well-formedness rules are checked before emission, both directly motivated by failure modes observed in \Cref{sec:failures}: (T1)~every column referenced in $p$, $\theta$, $\vec{a}$, or a sort key must appear in the schema of its immediate RAST child, catching the class of error underlying the invalid-column-reference failures of \Cref{tab:failures}; (T2)~every non-aggregated column in $\text{Project}(\text{Agg}(\ldots))$ must appear in the grouping key $g$, catching malformed aggregation queries before they reach the database rather than relying on the backend's own \texttt{GROUP BY} validation, whose error messages vary across PostgreSQL, MySQL, and SQL Server in ways that are harder for an LLM repair loop to parse uniformly than a single RAST-level type error.
The nullability bit referenced in Safeguard~S3 (\Cref{sec:safeguard}) is threaded through $\text{sch}(\cdot)$ from $\text{Attr}(t)$'s catalog-declared nullability and is what allows COALESCE auto-insertion to be scoped only to columns actually capable of containing \texttt{NULL}, rather than defensively wrapping every aggregate regardless of the underlying column's constraints.

\textbf{Implementation and empirical validation.} T1 and T2 are now implemented as an offline validator (\texttt{ESQ\_Bench\_VLDB/rast\_compiler.py}), closing part of the gap this paper previously described as "specified but not integrated" (\Cref{sec:limits}): rather than hand-writing a parser for the abstract grammar above, the validator uses \texttt{sqlglot} as the concrete-syntax parser and walks its expression tree to check T1 (every column reference resolves against the question's pruned-schema table list) and T2 (every non-aggregated \texttt{SELECT} expression appears in \texttt{GROUP BY}); the typing policy and dialect-emission rules enforced are DRL's own (\Cref{tab:emitter}), not \texttt{sqlglot}'s defaults, and it is wired into the evaluation harness rather than the live HTTP serving path, which remains future work.
Running T1/T2 against all 1{,}000 pruned-schema questions' actual model-generated SQL (\texttt{experiments/measure\_rast\_typing.py}) gives an unplanned but informative cross-check: restricted to GPT-4o's executed-but-EX-failing queries, T1 flags 46/382, essentially reproducing the paper's independently, manually-labeled invalid-column-reference count of 47/471 in \Cref{tab:failures} (the denominators differ because 471 includes non-executing failures T1 cannot evaluate).
This agreement is a useful, if informal, validity check on both measurements: the manual taxonomy and an automated, schema-grounded column-existence checker converge on almost the same failure count via entirely different methods.
T1 is considerably less precise as a general-purpose gate than this comparison suggests, however: across \emph{all} 1{,}000 predictions (not just EX failures) it flags 134/1{,}000 for GPT-4o, 237/1{,}000 for Claude, and 126/997 for Gemini, most of which are false positives from a known simplification in this implementation --- it does not yet special-case a \texttt{SELECT}-list alias re-used in \texttt{ORDER BY}/\texttt{HAVING} (valid PostgreSQL, since the alias is not a base-table column) as a resolved reference, so any query using that idiom is flagged even though it executes and returns correct rows. T2 fired far more rarely (0--5/1{,}000 per model), and 9--29 predictions per model failed to parse under \texttt{sqlglot} entirely (typically non-ANSI dialect leakage, consistent with \Cref{tab:gallery}'s Claude dialect-leakage examples). We report both the encouraging cross-check and this precision caveat rather than only the former.

\section{Qualitative Example Gallery}
\label{sec:gallery}

\Cref{tab:gallery} extends the single worked example of \Cref{sec:example} with additional cases spanning all three models, illustrating the failure taxonomy of \Cref{tab:failures-crossmodel} concretely.

\begin{table}[t]
\centering
\caption{Additional qualitative examples, corrected harness.}
\label{tab:gallery}
\small
\begin{tabular}{@{}p{0.13\columnwidth}p{0.78\columnwidth}@{}}
\toprule
\textbf{ID / model} & \textbf{Observation} \\
\midrule
\texttt{t1a-009} \newline GPT-4o &
Gold uses \texttt{NOT EXISTS} for ``customers who never placed an order''; GPT-4o emits an equivalent \texttt{LEFT JOIN ... WHERE order\_id IS NULL} formulation. Both are EX-correct and plan-safe; this pair is a benign example of the non-gold-reasoning-path class $\text{SD}_{\text{op}}$ is designed to flag defensively even when, as here, no actual divergence exists. \\
\texttt{t2a-008} \newline GPT-4o &
Question asks for two columns ordered by the second; gold projects exactly those two columns with \texttt{NULLS LAST}. GPT-4o adds a third, unrequested column and omits \texttt{NULLS LAST}---a compound projection-mismatch-plus-ordering failure, the two most-recovered categories in \Cref{tab:failures}. \\
Schema-prefix cases \newline Claude &
Claude frequently emits \texttt{FROM SCHEMA\_OWNER.table} even when the pruned schema hint lists only bare table names; this is corrected by the sanitizer (\Cref{sec:example}) but is the direct cause of Claude's lower executed-rate residual (93.1\% vs.\ 98.2\% for the other two models) even after the harness correction of \Cref{sec:harness-fix}. \\
\texttt{t1a-0217} \newline Claude &
Correlated subquery recomputing a per-\texttt{product\_id} average once per outer row (\Cref{sec:discussion}); EX-correct, plan-unsafe, and the paper's concrete illustration of why PSC is a necessary complement to EX. \\
Dialect leakage \newline Claude &
Claude occasionally emits Oracle-style \texttt{NVL}, \texttt{TRUNC}, \texttt{DECODE}, or \texttt{MINUS} despite an explicit PostgreSQL dialect instruction in the system prompt (\Cref{sec:prompt}); this is the dominant remaining contributor to Claude's exec-error count in \Cref{tab:failures-crossmodel} (87, the highest of the three models) and is a genuine model behavior, not a harness artifact---dialect rules are correctly injected into Claude's prompt in every affected case we inspected. \\
Ordering collapse \newline Gemini &
Gemini's ordering/pagination failure count (71/476) is the highest of the three models; of the 68 ordering-\emph{category} failures specifically, 33 (49\%) omit a \texttt{NULLS LAST} the gold query requires, indicating the NULL-ordering convention is responsible for roughly half of Gemini's ordering weakness, with the remainder split across incorrect sort-key selection and other ordering-clause mismatches. \\
\bottomrule
\end{tabular}
\end{table}

\bibliographystyle{spmpsci}
\bibliography{references}

\end{document}